\documentclass[prd,showpacs,nofootinbib,preprintnumbers]{revtex4}
\usepackage{latexsym}
\usepackage{amssymb}
\usepackage{amsmath}
\usepackage{subfigure}
\usepackage[dvips]{epsfig}
\usepackage{psfrag}
\usepackage{psfragx}
\usepackage{wrapfig}
\usepackage[dvips]{epsfig}
\numberwithin{equation}{section}

\newcommand{\bbn}{\bar{\bf n}}
\newcommand{\brmn}{\bar{\rm n}}

\newcommand{\lb}{\label}
\newcommand{\be}{\begin{equation}}
\newcommand{\ee}{\end{equation}}
\newcommand{\ba}{\begin{align}}
\newcommand{\ea}{\end{align}}
\newcommand{\bea}{\begin{eqnarray}}
\newcommand{\eea}{\end{eqnarray}}
\newcommand{\bw}{\begin{widetext}}
\newcommand{\ew}{\end{widetext}}

\newcommand{\e}{{\rm e}}

\newcommand{\nn}{\nonumber}
\newcommand{\zt}{\dot{z}}

\newcommand{\y}{\mathbf{y}}

\newcommand{\mm}{\mathrm{m}}
\newcommand{\n}{\mathbf{n}}
\newcommand{\Nv}{\mathbf{N}}
\newcommand{\NNv}{{\mathrm{N}}}

\newcommand{\pp}{\, ... \,}
\newcommand{\tot}{\mathrm{tot}}

\newcommand{\eff}{\mathrm{eff}}

\newcommand{\rad}{\mathrm{rad}}

\begin{document}

\title{\begin{flushright}\begin{small}    CCTP-2010-17
\end{small} \end{flushright} 
Classical ultrarelativistic bremsstrahlung in extra dimensions}

\author{Dmitry V. Gal'tsov$^*$, Georgios Kofinas$^\dagger$, Pavel Spirin$^{*\ddagger}$ and Theodore N.
Tomaras$^\dagger$\\
\vspace{0.3cm}
$^*$ Department of Theoretical Physics, Moscow State University,
119899, Moscow, RUSSIA\\
$^\dagger$ Department of Physics and Institute of Theoretical and Computational Physics,
 University of Crete, 71003 Heraklion, GREECE\\
$^\ddagger$ Bogolubov Laboratory of Theoretical Physics, JINR, Joliot-Curie 6, Dubna, RUSSIA\\
 \vspace{0.3cm}
{\tt E-mail: galtsov@phys.msu.ru, gkofin@phys.uoa.gr,
salotop@list.ru, \\
tomaras@physics.uoc.gr}}


\begin{abstract}
The emitted energy and the cross-section of classical scalar bremsstrahlung in massive particle
collisions in $D=4+d-$dimensional Minkowski space ${\mathcal M}_D$ as well
as in the brane world ${\mathcal M}_4\times {\mathcal T}^d$ is computed to leading ultra-relativistic order.
The particles are taken to interact in the first case via the exchange of a bulk massless scalar field $\Phi$
and in the second with an additional massless scalar $\phi$ confined together with the particles
on the brane. Energy is emitted as $\Phi$ radiation in the bulk and/or $\phi$ radiation on the
brane. In contrast to the quantum Born approximation, the classical result is unambiguous and valid in
a kinematical region which is also specified. For $D=4$ the results are in agreement with corresponding
expressions in classical electrodynamics.
\end{abstract}
\maketitle

\tableofcontents

\section{Introduction and results}
\label{intr}

There is little doubt that new physics will soon be discovered at the TeV scale.
Absence of a light fundamental higgs boson implies new physics and, furthermore,
a light higgs boson is unstable without new physics, of which examples are supersymmetry
broken at the TeV scale, or maybe some variant of gauge interaction
becoming strong at the TeV scale \cite{iliopoulos}.

A less well founded, but arguably more fascinating scenario for LHC would be the discovery of TeV
scale gravity and Large Extra Dimensions (LEDs) \cite{ADD}. Although extra dimensions were proposed
a century ago, their raison d' \^etre,
as well as their number, size and topology vary in different proposals. The potentially
most natural justification of the existence of extra dimensions with a definite proposal
for their number is string theory, while their most natural size is thought to be
the Planck length $\mathcal{O}(10^{-16}\; {\rm TeV}^{-1})$. Nevertheless, the idea that some
extra dimensions
could be of order $\mathcal{O}({\rm TeV}^{-1})$ appeared in the context of
heterotic string theory, when the size of extra dimensions was connected to the supersymmetry
breaking scale \cite{ablt}.
Earlier thoughts of the Universe as a topological defect embedded in a higher-dimensional space-time
\cite{defect} and the development of the Brane-World idea led to a variety of alternative scenaria
\cite{RS,UED} reviewed in \cite{reviews}.
They differ in the topology, size and geometry of the extra dimensions, as well as
in the nature of the degrees of freedom which may propagate in the bulk.
The technically simplest such model is the ADD \cite{ADD,GRW},
with the Standard Model particles confined on the brane and only gravity allowed to propagate in
the whole ${\mathcal M}_4\times {\mathcal T}^d$ space-time.

The invention of ways to detect and measure the size of extra dimensions has attracted
considerable attention and the study of radiation is an obvious candidate.
Indeed, it has been argued that in the ADD model graviton bremsstrahlung from neutron
collisions may give significant contribution to the cooling of supernovae \cite{bremnr},
thus providing strong bounds on relevant parameters. Also, production of
Kaluza-Klein (KK) gravitons in collisions of relativistic
particles at LEP or LHC was extensively discussed in the literature, with emphasis on processes
efficient to at least demonstrate their very existence \cite{MPP,Dvergsnes:2002nc}. However,
the actual quantum perturbative calculations in the context of ADD are ambiguous, as they
are plugged with tree-level divergences, associated with the emission of infinite
momentum of modes into the compact dimensions \cite{GRW}. This problem is
manifest already in the calculation of any elastic scattering
amplitude in such spaces and was treated in rather ad-hoc ways.
Contrary to the above situation, a purely classical computation, which by itself carries a natural
mechanism to cut-off the energetic KK excitations, leads to unambiguous results \cite{gkst}.
Theoretically, such an approach is justified
by the fact that at transplanckian energies scattering processes are dominated by gravity,
which in addition, may be treated classically, at least in some range of
momentum transfer \cite{thooft}. Furthermore, the classical formula for
the elastic scattering in ADD was shown to coincide with the one obtained in the
non-perturbative eikonal approximation, given by the summation
of an infinite number of ladder graphs in quantum theory \cite{gkst}.

Encouraged by the above result, one may try to deal in the same way
with the phenomenon of ultra-relativistic gravitational
bremsstrahlung, and the present paper is a first step in this
direction. To further motivate this approach, it is well known that
in 4-dimensions the classical treatment of relativistic
bremsstrahlung corresponds to non-perturbative calculations in
quantum electrodynamics. Specifically, bremsstrahlung in
ultra-relativistic scattering off a Coulomb center has been computed
using the exact Coulomb solutions of the Dirac equation (see
\cite{bere} and references therein). In the ultra-relativistic limit
and with momentum transfer not exceeding the particle mass, a good
approximation to the corresponding cross-section is given by the
simplified Furry-Sommerfeld-Maue formula, which for photon energy
much smaller than $E^2/m$, where $E$ and $m$ are the particle energy
and mass respectively, reduces to the classical result
\cite{GaGra}. The result is clearly non-perturbative in the normal
sense of expansion in powers of the fine structure constant. No
analogous non-perturbative quantum calculations are available for
bremsstrahlung in higher dimensions, especially in the presence of
compact LEDs. Although certain aspects of classical radiation in
$D-$dimensional Minkowski space were discussed recently
\cite{minkd,minkd1}, radiation in the presence of compact LEDs has
not been considered so far. So, the hereby proposed classical
treatment, which as explained in \cite{gkst} is unambiguous and
expected to be trustable in the ultra-relativistic limit and small
scattering angle, is also, to the best of our knowledge, the first
actual computation of the phenomenon in higher dimensional
space-times with the topology of ${\mathcal M}_4\times{\mathcal
T}^d$.

The purpose of this paper is to present a classical computation of bremsstrahlung in particle collisions
on a flat 3-brane embedded in flat space-time of arbitrary dimension and with the transverse
space taken either euclidean $\mathcal{E}^d$ (relevant to the scenario of Universal Extra Dimensions
\cite{UED})
or toroidal ${\mathcal T}^d$ (as in ADD) with equal radii.
As a first step, the bulk interaction will be {\it modeled} by the exchange of a massless
scalar field $\Phi$, while an additional massless scalar $\phi$, confined on the
brane, will be introduced to mimic, in the case of toroidal extra dimensions, the Standard Model interactions
between the colliding particles. The case of gravity differs, of course, crucially from massless scalar
exchange. Nevertheless, the present analysis captures several important {\it technical} points
of gravitational bremsstrahlung \cite{gkst3}, to be presented in full detail elsewhere \cite{gkst4}.

Straightforward classical perturbation theory, developed in
Section II, is applied to iteratively solve the particle equations
of motion and the field equations as well. The {\it radiation
efficiency} $\varepsilon$, i.e. the fraction of the initial energy
emitted as bremsstrahlung radiation, is computed in Section II to
leading ultra-relativistic order and low momentum transfer and is
found to be \be \varepsilon_d \equiv \frac{E_{\rad}}{E}\simeq
{\mathcal C}_d\, (\gamma\, r_d^3 / b^3)^{1+d}\, , \label{varepsilonD}
\ee where $r_d$ is the $D-$dimensional classical radius of the
colliding particles and $b$ the impact parameter. The constant
${\mathcal C}_d$ is given in (\ref{CD}).

The case of toroidal extra dimensions is dealt with in Section III.
The light $\Phi-$KK modes participate crucially both in the interaction between
the colliding particles and in the radiation emitted in the bulk. These roles are studied
separately with appropriate choices of their couplings to the radiating particles.
The radiation efficiency, depending on which interaction is assumed to be dominant
(via $\Phi$ or $\phi$) and on the nature of radiation ($\Phi-$radiation in the bulk or $\phi$ on the brane),
is to leading ultra-relativistic order shown in Table \ref{tableM4Td}.
Note the interrelations among the various entries of the Table as well as with
(\ref{varepsilonD}). The range of validity of the computation, the consequences of the results
and the modifications expected in actual gravity are presented in the discussion Section IV \cite{gkst4}.
Finally, three Appendices in the end contain derivations of formulas used and
clarifications of technical approximations made in the text. They also contain a qualitative explanation
of the basic features of Table I.
\begin{table}[ht]
\vspace{0.3cm}
\centering
\begin{tabular}{||l||c|c||}
\hline \hline
\;\;\;\;\;\;\;\;\;\;\;\;\;\;\;\;\;\;\; Radiation &Bulk ${\mathcal M}_4\times {\mathcal T}^d$& Brane ${\mathcal M}_4$ \\
Interaction & & \\
 \hline
 \hline
  &  &  \\
 Bulk ${\mathcal M}_4\times {\mathcal T}^d$ & $\;\;\;\; \;\;\;
\varepsilon_{\Phi\Phi}
\simeq {\mathcal C}_d\, (\gamma\, r_d^3 / b^3 )^{1+d}
 \;\;\;\;\;\;\;$ &
 $\;\;\;\;\;\; \varepsilon_{\Phi\phi}\simeq
{\mathcal C}_{\Phi\phi}\, \gamma \, \frac{R_0}{b}\, (\frac{r_d}{b})^{2(1+d)}  \;\;\;\;\;\; $ \\
  &               &  \\
 \hline
  &  &  \\
 Brane ${\mathcal M}_4$ & $\varepsilon_{\phi\Phi}
\simeq  {\mathcal C}_{\phi\Phi}\, \left(\frac{R_0}{b}\right)^2 \left(\gamma\,\frac{r_d}{b}\right)^{1+d}$
& $\;\;\;\varepsilon_{\phi\phi} \simeq {\mathcal C}_0 \, \gamma \, R_0^3 / b^3 \;\;\;
$ \\
  &   &  \\
   \hline \hline
 \end{tabular}
\caption{Radiation emission efficiency in ${\mathcal M}_4\times {\mathcal T}^d$. The standard
``electromagnetic" ($R_0$)
and the $D-$dimensional ``gravitational" ($r_d$) classical radii of the colliding particles, as well as the
constants ${\mathcal C}_{ab}$ and ${\mathcal C}_0$ are given in the text.}
\label{tableM4Td}
\end{table}


\section{Scalar ultra-relativistic bremsstrahlung in ${\mathcal M}_D$}
\label{sc_sc}
In this section we formulate our approach in the simplest case of
two massive \footnote{The interest here is in strictly massive particles.
Proper treatment of the massless particle case would require the use of the Polyakov
action for them.} scalar point charges interacting via a linear scalar field in
Minkowski space of arbitrary dimension $D$. This will serve as a
reference calculation for the subsequent treatment of compactified extra dimensions.

\subsection{The model - Generalities}

Consider two particles with masses $m$ and $m'$ moving along the
world-lines $x^M=z^M(\tau), \,x^M=z'^M(\tau')$ interacting
with a massless scalar field $\Phi$ with coupling constants $f$ and
$f'$, respectively. The relevant action is
 \begin{equation}
 \label{action0}
S=-\int d\tau \sqrt{\zt^2}[m+f \Phi(z)] -\int d\tau'
\sqrt{\dot{z}^{'2}}[m'+f' \Phi(z')]+\frac{1}{2}\int d^D x
\partial_M\Phi\,
\partial^M\Phi,\,
 \end{equation} with $M, A, B, \ldots = 0,1, \ldots ,D-1=3+d$, and the
metric signature is $(+,-,-,\ldots ,-)$. $\dot{z}^2\equiv
\eta_{MN}\dot{z}^M\dot{z}^N$, and the dot denotes differentiation
with respect to $\tau$ or $\tau'$. The classical radius $r_d$ of
the particle $m$ is defined by
\be
\frac{f^2}{m} \equiv r_d^{d+1}\, ,
\label{rd} \ee
and similarly for $m'$.

Linearity of the field equations implies that
$\Phi=\Phi_m+\Phi_{m'}$, where
\begin{equation}
\label{phim}
\Box\, \Phi_m(x) = \rho (x),\quad \Box\, \Phi_{m'}=\rho' (x),\quad \Box\equiv
\eta^{MN}\partial_M\partial_N,
\end{equation}
the sources being
 \begin{equation}\label{rhophi}\rho (x)= f\int d\tau\sqrt{\dot{z}^2}
\delta^D(x-z(\tau)),\quad \rho' (x)=f'\int d\tau'
\sqrt{\dot{z}'^2} \delta^D(x-z'(\tau')).
\end{equation}
The particles' equations of motion read
\begin{equation}
\frac{d}{d\tau}\left[(m+f\Phi)\dot{z}^M\right]=f\partial^M \Phi,
\end{equation}
and similarly for $m'$, provided the parameters $\tau, \tau'$ are chosen so that
\begin{equation}
\label{gauge}
\dot{z}(\tau)^2=1=\dot{z}'(\tau')^2.
\end{equation}
This can be rewritten as
\begin{equation}
(m+f\Phi)\ddot{z}^M=f \, \Pi^{MN}\partial_N \Phi,\quad
\Pi^{MN}=\eta^{MN}-\dot{z}^M \dot{z}^N,
\end{equation}
where $\Pi^{MN}$ is the projector onto the space orthogonal to the world-line.
In this equation the field on the right hand side
contains both the self-action term ($\Phi_m$), and the mutual
interaction term $\Phi_{m'}$. The first gives rise to divergences,
taken care of by mass renormalization, and also for
$D\geqslant 4$ to the introduction of higher derivative classical
counter-terms \cite{minkd}, which for simplicity will be ignored. Thus, the equations of motion become:
\begin{equation}
\label{eoms}
(m+f\Phi_m)\ddot{z}^M=f\Pi^{MN}\partial_N \Phi_{m'},\quad
(m'+f'\Phi_{m'})\ddot{z}'^M=f' \,\Pi'^{MN}\partial_N \Phi_m.
\end{equation}

Fourier transform using
$\Psi(x)\equiv \int \Psi (k)\e^{-i k\cdot x} d^D k/(2\pi)^D$ and
$\Psi(k)\equiv \int \Psi (x) \e^{i k\cdot x} d^D x$. The
retarded solutions of the wave equations (\ref{phim}) can then be written
algebraically:
\begin{align}
\label{retardedsolution}
\Phi(k)= G_{\rm ret}(k)\rho(k),\quad G_{\rm ret}(k)=\bar{G}(k)+ i
\pi \varepsilon (k^0)\delta(k^2),\quad \bar{G}(k)=- \; \mathcal
P\frac{{1}}{k^2},
\end{align}
with the Fourier-transforms of the source terms given by:
\begin{equation}
\label{rhophi(k)} \rho (k)=f \int   \e^{ik\cdot z(\tau)}d\tau ,\quad
\rho' (k)=f' \int \e^{ik\cdot z'(\tau')}d\tau' .
\end{equation}

The energy-momentum $P^M$ of the scalar radiation, emitted by the
particles during the scattering process, can be computed by
integrating the field stress-tensor
\begin{equation}
\label{tabphi}
T^{MN}= \partial^M\Phi \,
\partial^N \Phi-\frac{1}{2}\eta^{MN}  \partial_L\Phi \,
\partial^L \Phi
\end{equation}
between the space-like hypersurfaces $\Sigma_{\pm\infty}$
corresponding to $t\to\pm\infty$ and reads:
\begin{equation}
\label{momecha}
P^M=\int\limits_{\Sigma_{\infty}} T^{MN} dS_N-
\int\limits_{\Sigma_{-\infty}} T^{MN} dS_N.
\end{equation}
 Given the fall-off of the stress-tensor at spatial infinity
 $T^{MN} \sim 1/r^{D-2}+\mathcal{O}(r^{-(D-1)})$
 (derived from the behavior of the Li\'enard-Wiechert potentials), one can
write $P^M$ as an integral over the closed boundary of the
space-time tube and, subsequently, transform it to  the volume
integral $P^M= \int d^D x \nabla_N T^{MN}$, which upon substitution of
(\ref{tabphi}) leads to
\begin{equation}
P^M=\int d^Dx (\rho+\rho') \;\partial^M \Phi,
\end{equation}
or equivalently
\begin{align}
P^M=-\frac{i}{(2 \pi)^D}    \int d^D k \,k^M\,\Phi (k)
\left[\rho^* (k)+ \rho'^* (k)\right].
\end{align}
Substituting here the retarded solution (\ref{retardedsolution})
 and taking into account that the contribution of
the principal part $\bar{G}(k)=- \; {\mathcal P}\, (k^2)^{-1}$
vanishes being odd under $k^M \to -k^M$, one is led to
\begin{align}
\label{pa}
P^M=\frac{1}{(2 \pi)^{D-1}}
 \int d^D k \, \theta(k^0)
 k^M  \delta(k^2) |\rho (k)+\rho' (k)|^2.
\end{align}
Finally, denote the frequency $k^0\equiv \omega$, use
$\theta(k^0)\delta(k^2)=\delta(\omega-|\mathbf{k}|)/2|\mathbf{k}|$,
and integrate over $|\mathbf{k}|$ to obtain the
spectral-angular distribution of radiation:
\begin{align}
\label{pafinal}
\frac{dP^M}{d\omega d\Omega}= \frac{1}{2(2 \pi)^{D-1}}
\omega^{D-3}  k^M  |\rho (k)+\rho' (k)|^2
\Big|_{k^M=\omega (1,\mathbf{n})},
\end{align}
where $\Omega$ or $\mathbf{n}$ parametrize
the unit sphere in the $(D-1)-$dimensional Euclidean subspace.

\subsection{Perturbative solution}

\subsubsection{Particle trajectories}

We next solve the equations of motion for the particles and the field
$\Phi$ iteratively in powers of the coupling constants
$f$ and $f'$. For the particle trajectories we write
\begin{equation}
z^M(\tau)=b^M+u^M \tau +\delta z^M(\tau),\quad z'^M(\tau)=u'^M
\tau +\delta z'^M(\tau),
\end{equation}
(assuming $z^M(0)=b^M, z'^M(0)=0$) where $u^M$ and $u'^M$ are the
unperturbed constant four-velocities, specified by the initial
condition and chosen to satisfy $u^2=1=u'^{2}$. Combine with
(\ref{gauge}) to conclude that the perturbations of velocities
must be  orthogonal to the unperturbed world lines:
\begin{equation}
\label{orthogonality}
u_A\,\frac{d}{d\tau}(\delta z^A)=0\; ,\qquad  u'_A\,\frac{d}{d\tau'}(\delta{z'}^A)=0\;.
\end{equation}
To zeroth order, the solution of the wave equation describes
the non-radiative Coulomb field of two non-interacting particles,
which in terms of the Fourier-transform reads
\begin{align}
\label{phi(1)}
\Phi_m(q)= -\frac{2\pi f}{q^2}   \e^{iq\cdot z(0)}\delta(q\cdot u).
\end{align}
 Substituting this into the equations of motion (\ref{eoms}) we find
for  the first order perturbation:
\begin{equation}
\frac{d^2}{d\tau^2}{z}^A(\tau)=\frac{f}{m}\Pi^{AB}
\partial_B\Phi_{m'}(x)\Big|_{x^A=z^A(\tau)},
\end{equation}
where $\Pi^{AB}=\eta^{AB}-u^A u^B$
 is the projector on its unperturbed world-line. The solution for $\delta z$ is
\begin{equation}
\label{z2} \delta z^M(\tau)=- i \frac{f f'}{m(2\pi)^{D-1}} \int
\frac{d^D q}{q^2 (q\cdot u)^2} \e^{-iq\cdot b} \left(\e^{-i(q\cdot
u)\tau}-1+iq\cdot u\tau\right) \delta(q\cdot u') \Pi^{MN} q_N\,,
\end{equation}
with an analogous expression for $\delta z'$ and with the exchange
$u \leftrightarrow u', b \to -b$. They automatically satisfy the
orthogonality conditions (\ref{orthogonality}).

\subsubsection{The radiation and its main features}

The next step is to compute corrections to the field $\Phi$ due to
these perturbations of the trajectories. These will correspond to
the lowest order radiation field. To compute radiation using
(\ref{pafinal}) one needs the Fourier transform of the
corresponding sources. Our treatment is symmetric in
$m$ and $m'$, so we write only unprimed quantities. The zeroth
order terms do not contribute to radiation, since the
corresponding Fourier transforms (\ref{phi(1)}) vanish on the
light cone $k^2=0$ (except for the trivial point $k^\mu=0$). Ditto for the last two terms inside the parenthesis
in the integrand of (\ref{z2}). Thus, to lowest
non-trivial order \footnote{A word of caution about notation: To avoid the introduction
of too many symbols, we use the same symbol $\rho$ (a) for the generic source in the
previous section, (b) for its zeroth-order and first-order expressions, (c) for the sum of the
two that one has to substitute in the energy loss formula (\ref{pa}), with the zeroth order term
giving zero contribution. Hopefully, the context makes this notational simplification
harmless.}:
\begin{equation}
\rho(k) \simeq i  f \int d\tau \,\e^{ik\cdot z(0)+i(k\cdot u)\tau}(k\cdot \delta z(\tau)).
\end{equation}
Substitution of (\ref{z2}) gives
\begin{equation}
\label{rho3+3}
\rho(k)\simeq \frac{ f^2 f'}{(2\pi)^{D-2}} \frac{\Pi_{MN}k^M I^N \e^{ik \cdot z(0)}}{m(k\cdot u)^2},
\end{equation}
with
\begin{equation}
\label{I} I^M(k)\equiv \int d^D q \frac{\delta(q\cdot
u')\delta((k-q)\cdot u)\,  \e^{-iq\cdot b}}{q^2}q^M.
\end{equation}
The corresponding integral $I'_A(k)$ for the partner particle is
obtained from $I_A$ by the substitution $u\leftrightarrow u',
z_A(0)\leftrightarrow z'_A(0)$, $b_A \to -b_A$. The integrals are
evaluated in Appendix I in a fully covariant form.

It is convenient to work in the {\it lab frame}, in which $u'^{A}=(1,0,0,...,0)$,
i.e. with $m'$ initially at rest. Define the remaining
kinematical variables in this frame by
 \begin{equation}
 u^A=\gamma(1,{\bf{v}}),\;\;
 k^A=\omega(1,{\bf{n}}),\;\;
k\cdot u=\gamma\omega\xi,\;\;
\xi\equiv 1-v\cos\theta,\;\; k\cdot u'=\omega,
 \end{equation}
where as usual $\gamma=1/\sqrt{1-v^2}$ and $\theta$ is the angle between
 $\bf{n}$  and $\bf{v}$. The invariant expression for the Lorentz factor is $\gamma =u\cdot u'$.

Furthermore, we define the zeroes of the parameters $\tau$ and
$\tau'$ in such a way that at the point of closest proximity of
the scattered particles they satisfy $u\cdot b=0={u'}\cdot b$. In
other words, the``initial conditions" defined at $\tau=0=\tau'$
are given at the moment of least distance between the particles.
But then, $b^0=0$ and $\mathbf{v}\cdot \mathbf{b}=0$ (easiest
visualized in the CM frame).  The vectors $\bf{v}$ and
$\mathbf{b}$ define the collision plane, while the generic
$(D-1)$-dimensional unit vector ${\bf n}$ can be decomposed as:
\begin{equation}
\bf{n}=\frac{\bf{v}}{|\bf{v}|} \cos\theta+
\frac{\mathbf{b}}{|\mathbf{b}|}\sin\theta\sin\psi+\bf{m}\sin\theta\cos\psi,
\end{equation}
where $\bf{m}$ is a $D-1$ dimensional unit vector orthogonal to
the collision plane. Then
\begin{equation}
{\bf{k}}\cdot {\mathbf{b}}=-k\cdot b=b\,\omega\sin\theta \sin\psi,
\end{equation}
where the impact parameter $b\equiv |\mathbf{\Delta}|$ is also given by the invariant expression:
\begin{equation}
b=|\mathbf{b}| =\left(-b^2-\frac{[(u'\cdot b)u-(u\cdot
b)u']^2}{(u\cdot u')^2 v^2} \right)^{1/2}.
 \end{equation}

Occasionally, it will be convenient to express the angular dependence of
radiation in the rest frame of the moving particle $m$. In that
frame the radiation angle will be denoted by $\theta_0$, which is related to
$\theta$ via
\begin{align}
\label{hhh26a}
\cos \theta_0 =\frac{1}{v}\left(\frac{1}{\xi \gamma^2}
-1\right), \quad \sin \theta_0=\frac{\sin \theta}{\xi\gamma}.
 \end{align}

Using formula (\ref{Iaspecial}) of Appendix I one obtains
\begin{equation}
\label{rho3}
\rho(k)\simeq \frac{f^2f' e^{ik\cdot
z(0)}}{(2\pi)^{n+1} m\gamma^3 v^3 b^{2n} z \xi} \left[(\gamma^2\xi
-1) z \hat{K}_n(z)+ i v^2
\sin\theta\sin\psi\hat{K}_{n+1}(z)\right],
\end{equation}
where $n\equiv D/2 - 2 \equiv d/2$. The ``hatted'' functions $\hat{K}_n(z)\equiv z^n K_n(z)$.
Their invariant argument is
\begin{equation}
z=\frac{(k\cdot u)\, b}{\gamma v}=\frac{b\, \omega\, \xi}{v}.
\end{equation}

Similarly, the leading contribution of $m'$ to the radiation field in the lab frame is
\begin{equation}
\label{rho'3}
\rho'(k)\simeq -\frac{f'^2f}{(2\pi)^{n+1}m'\gamma^2 v^2
b^{2n}}\frac{\e^{ik\cdot z'(0)}}{z'}
 \left[\gamma\cos\theta z' \hat{K}_n(z')+
 i\sin\theta\sin\psi\hat{K}_{n+1}(z')\right],
 \end{equation}
 with
 \begin{equation}
 z'=\frac{(k\cdot u')b}{\gamma v}=\frac{b\, \omega}{\gamma v}.
 \end{equation}
Substitution in (\ref{pafinal}) gives the angular and frequency distribution of the emitted radiation.

\subsubsection{Frequency and angular distribution.}

(a) {\it The low-frequency limit.} Using
$\displaystyle \lim_{z\to 0}\hat{K}_n(z)=2^{n-1}\Gamma(n)$ for $ n
> 0$  and $ \displaystyle  \lim_{z\to 0} K_0(z) \simeq \ln(2/z) - C_E$
(where $C_E$ is the Euler constant) in (\ref{rho3}) one concludes that for $\omega\to 0$
the leading term for any $D$ will be the second term in
(\ref{rho3}) and
 \begin{equation}
 \rho(k)\left. \vphantom{\sqrt[n]{d}}\right|_{\omega \to 0}\simeq -\frac{i
f^2f'\Gamma(n+1)}{(2\pi)^{n+1}m\gamma^3 b^{2n+1}}
\frac{\sin\theta\sin\psi}{ (1-v\cos\theta)^2}\,\frac1{\omega
}\,.
\lb{lowfrequency}
\end{equation}
The divergence in $\omega$, is reminiscent of the infrared
divergence of the corresponding Feynman  diagrams. However, upon
multiplication by $\omega^{D-3}$ from the measure and integration,
it contributes a finite amount to the radiation loss. Furthermore,
the non-zero value of $dE_{\rad}/d\omega(\omega=0)$ for $D=4$ is
tiny for large $\gamma$.

\vspace{0.5cm}

(b) {\it Frequency cut-off and beaming.} The exponential decay of
$\rho(k)$ due to $\displaystyle \lim_{z\to\infty}\hat{K}_n(z)\sim
\sqrt{\frac{\pi}2} z^{n-1/2}\e^{-z}$ gives a $\theta-$dependent
upper cut-off for the frequency of the emitted radiation:
\begin{equation}
\omega <
\omega_{\rm cr}(\theta)= \frac v{b(1-v\cos\theta)}.
 \end{equation}
In the ultrarelativistic case
 \begin{equation}
 \label{mac-salo}
 \omega_{\rm cr}(\theta)\simeq \frac2{b(\theta^2+\gamma^{-2})},
 \end{equation}
so that most of the radiation is beamed inside the cone \be \theta
< \gamma^{-1} \lb{theta} \ee in any dimension $D$. The maximum
frequency of radiation is
 \begin{equation}
 \label{omax}
 \omega_{\rm max}\simeq \omega_{\rm cr}\Big|_{\theta=0} \sim
\frac{2\gamma^2}{b}.
 \end{equation}

On the other hand, $\rho'(k)$ does not exhibit sharp anisotropy in $\theta$ and the associated
frequency cut-off in the ultra-relativistic case is $2\gamma$ times smaller than $\omega_{\rm
max}$:
 \begin{equation}\omega'_{\rm max}\sim \frac{\gamma}{b} \sim \frac{\omega_{\rm
max}}{2\gamma}.
 \end{equation}

\begin{figure}
\begin{center}
\includegraphics[angle=0,width=10cm]{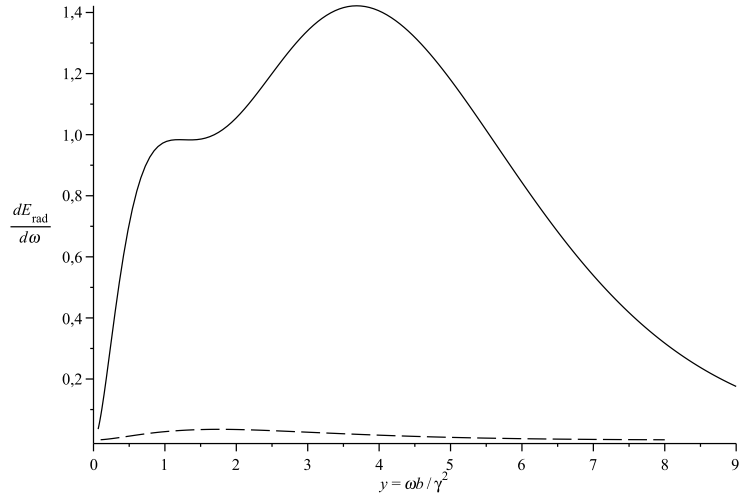}
\caption{The contribution to the frequency distribution of the
emitted energy of the real (solid line) and the imaginary (dashed
line) part of $\rho$ for $D=6$ and $\gamma=20$. The two curves are
rescaled by the same factor.} \label{real_vs_imaginary_rho}
\end{center}
\end{figure}

\vspace{0.5cm}
(c) {\it The imaginary part of $\rho(k)$ is negligible.} One might expect, that as in $D=4$ electrodynamics
one can combine the low-frequency
amplitude (\ref{lowfrequency}) with this cut-off to estimate the total radiation loss. This turns out to
be incorrect here. The imaginary part of $\rho$ in (\ref{rho3}) is suppressed compared to its real part by
a factor $\sin\theta\sim \theta<1/\gamma$. So, the leading ultrarelativistic contribution to the radiation
loss is due to the real part of $\rho$, i.e. the first term of the amplitude (\ref{rho3}). This is demonstrated
for $D=6$ and $\gamma=20$ in Figure \ref{real_vs_imaginary_rho}.

\begin{figure}
\begin{center}
\includegraphics[angle=0,width=10cm]{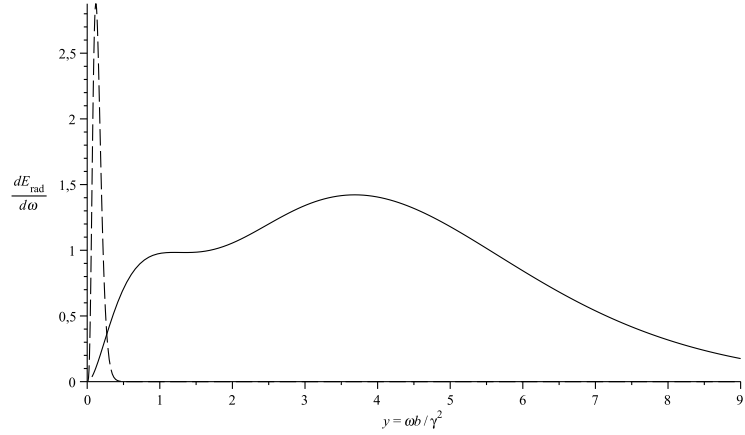}
\caption{The relative contribution to the frequency distribution of the emitted energy of $\rho(k)$
(solid line) and $\rho'(k)$ (dashed line) for $D=6$ and $\gamma=20$. The two curves are
rescaled by the same factor.}
\label{primed_vs_unprimed}
\end{center}
\end{figure}

\vspace{0.5cm}
(d) {\it $\rho'(k)$ is also negligible.} Using (\ref{intfreq}) and (\ref{jjj6}) one may
convince oneself that in the lab frame
the contribution to the total radiation (a) of the particle $m'$ and
(b) of the cross-term in (\ref{pafinal}) are both negligible in the ultra-relativistic limit.
In particular, Figure \ref{primed_vs_unprimed} shows
separately the contributions of $m$ (solid line) and $m'$
(dashed line) to the emitted energy in the lab frame. So, the radiation from $m'$ is important at
very low frequencies,
but may be neglected everywhere else, as well as in the total energy loss.
Thus, we shall neglect the contribution of everything else but the real part of $\rho(k)$. This
gives most of the energy lost in the collision.

\vspace{0.5cm}
(e) Finally, it is convenient to introduce the angle $\hat\theta$ by
 \begin{equation}
 \cos{\hat\theta} \equiv v,\quad \sin\hat{\theta}= 1/\gamma.
 \end{equation}
 Within the cone $0\leqslant \theta \leqslant \hat{\theta}$ the quantity $\xi$ for
ultrarelativistic  velocities remains small varying in the region
$1/(2\gamma^2) \leqslant  \xi \leqslant  1/\gamma^2$. Note that for $\theta=\hat\theta$
the $m$-rest-frame radiation angle is $\theta_0=\pi/2$.

\vspace{0.5cm} (f) {\it The angular distribution
$dE_{\rad}/d\theta$.} As argued above, the leading contribution
{\it to the total energy loss} is due to the real part of the
amplitude $\rho(k)$, i.e.
\begin{equation}
\label{rho_u0}
\rho(k)\simeq {\rm Re}\;\rho (k)\simeq -\frac{f^2f'}{(2\pi)^{n+1}m v^3 \gamma^3 b^{2n}}
\left(\frac{1}{1-v\cos\theta}-\gamma^2\right)\; \hat{K}_n(z)\, ,
\end{equation}
which does not depend on angles other than the main polar angle $\theta.$

\begin{figure}
\centerline{
\includegraphics
[angle=0,width=12cm]{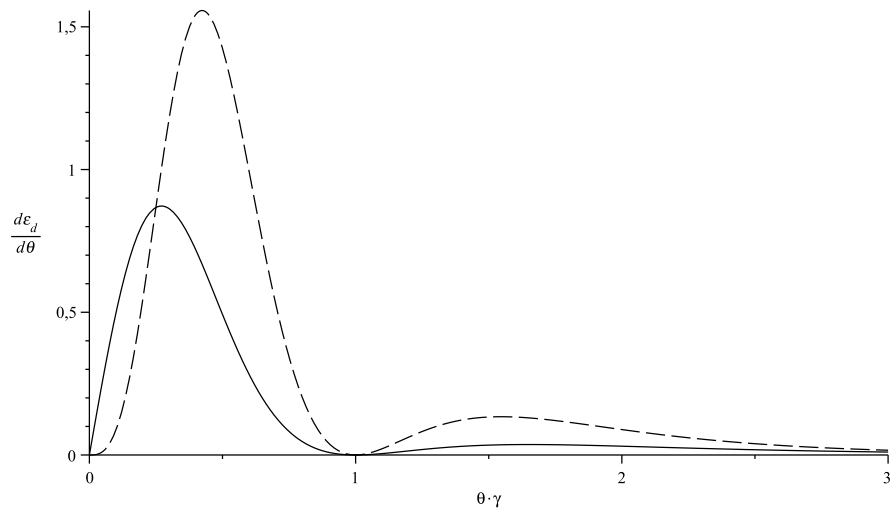}}\caption{The angular distribution
$d\varepsilon_d/d\theta$ versus $\theta/\hat\theta$ for
$\gamma=500$, $r_d/b=0.3$ and for $d=0$ (solid line) and $d=2$
(dashed). The maxima are at $\theta_{1,2}\simeq \left[\frac{7+d\mp
2\sqrt{11+2d}}{5+d}\right]^{1/2} \hat{\theta}.$} \label{theta_distr}
\end{figure}

Substitute (\ref{rho_u0}) into (\ref{pafinal}) and use (\ref{intfreq}) to integrate over the
frequencies. The result is the angular
distribution of radiated energy:
\begin{align}
\label{angl_distr} \frac{d  E_{\rad}}{d\Omega}= \frac{\sqrt{\pi}\,
\Gamma\left(\frac{3D-9}{2} \right) \Gamma
\left(\frac{2D-5}{2}\right)\Gamma \left(\frac{D-1}{2} \right)}{
8(2 \pi)^{2D-3} \Gamma(D-2)} \frac{f^4 f'^2}{b^{3D-9}\gamma^6
m^2}\,\xi^{-(D+1)}\,\left(1-\gamma^2 \xi\right)^2\,.
\end{align}
Integration over $\psi$ gives to leading ultra-relativistic order the $\theta-$distribution
\begin{align}
\label{dEdtheta}
\frac{d E_{\rm rad}}{d \theta}\simeq \frac{
2^{4-3D} \Gamma\left(\frac{3D-9}{2} \right) \Gamma
\left(\frac{2D-5}{2}\right)}{\pi^{3(D/2-1)} \Gamma^2
\left(\frac{D-2}{2} \right)} \frac{f^4
f'^2}{b^{3D-9}\gamma^6 m^2}\,\xi^{-(D+1)}\,\left(1-\gamma^2 \xi\right)^2\,\sin^{D-3}\theta\,.
\end{align}
In the special case $f=f'$ and $m=m'$ one obtains for the radiation {\it efficiency}
$\varepsilon_d\equiv E_{\rm rad}/E$
\begin{align}
\label{depsilondtheta} \frac{d\varepsilon_d}{d\theta} \simeq
\frac{\Gamma\left(\frac{3+3d}{2} \right)
\Gamma\left(\frac{3+2d}{2}\right)}{2^{8+3d}\,\pi^{3+3d/2}
\Gamma^2\left(1+d/2 \right)}\,
\gamma^{-7}\left(\frac{r_d}{b}\right)^{3(1+d)} \xi^{-5-d} \left(
1-\gamma^2 \xi\right)^2 \,\sin^{1+d}\theta \,,
\end{align}
plotted for various dimensions, $\gamma=500$ and $r_d/b=0.3$ in Figure \ref{theta_distr}.

\vspace{0.5cm}

\begin{figure}
\centerline{
\includegraphics
[angle=0,width=12cm]{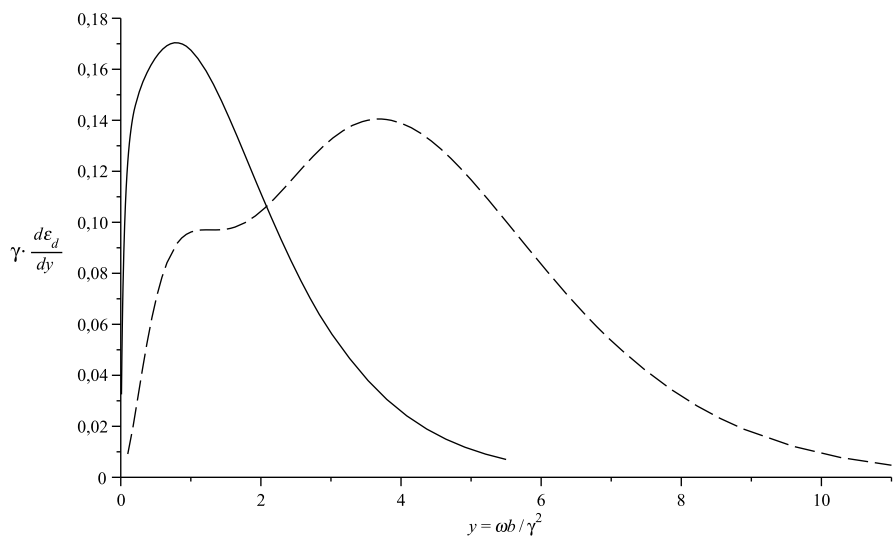}}\caption{Frequency distribution
$d\varepsilon_d/ dy$ versus $y\equiv b\,\omega/\gamma^2$ for
$\gamma=500$, $r_d/b=0.3$ and for $d=0$ (solid curve) and $d=2$
(dashed line). \label{omega_distr}}
\end{figure}

(g) {\it The frequency distribution $dE_{\rm rad}/d\omega$.}
Similarly, define $y\equiv \omega\, b/\gamma^2$ and substitute (\ref{rho_u0}) into (\ref{pafinal}) to obtain
\begin{align}
\label{angl_theta_distr} \frac{d \varepsilon_d}{d y}=
  \frac{\pi^{1+d/2}}{(2\pi)^{5+2d}\Gamma \left(\frac{2+d}{2}
\right)} \gamma^{3+2d} \left(\frac{r_d}{b}\right)^{3(1+d)} y^{2+d}
\!\! \int \! d \theta \left[1-(\gamma^2\xi)^{-1}\right]^2
\hat{K}_{d/2}^2(y \gamma^2 \xi) \sin^{1+d} \theta \,,
\end{align}
evaluated numerically and
shown for various dimensions, $\gamma=500$ and
$r_d/b=0.3$  in
Figure \ref{omega_distr}.

\subsubsection{The total radiated energy $E_{\rad}$}

Finally, using (\ref{jj2q}) one may integrate (\ref{dEdtheta}) over
$\theta$ to obtain the leading ultra-relativistic limit of the
radiation efficiency (for $m=m'$ and $f=f'$)
 \be \label{sc12}
\varepsilon_d \equiv \frac{E_{\rad}}{E}\simeq {\mathcal C}_d
\left(\gamma\, \frac{r_d^3}{b^3}\right)^{1+d},
 \ee with \be
{\mathcal C}_d\equiv \frac{ (2+d) \,\Gamma \left(\frac{3+3d}{2}
\right) \Gamma \left(\frac{3+2d}{2} \right)}{2^{7+2d}
\pi^{3(2+d)/2}\Gamma(4+d)}. \lb{CD} \ee The radiation loss grows
with $d$ as $\gamma^{2+d}$.
For $d=0$
$$
\varepsilon_0 \simeq \frac{1}{3\times
2^9\, \pi^2}\; \gamma \, \left(\frac{r_0}{b}\right)^3
$$
in agreement (modulo polarizations) with the corresponding formula in
classical electrodynamics \cite{LandauII}.

\subsubsection{The bremsstrahlung cross-section $\sigma_D$}

The energy differential cross-section with
dimension (length)$^{2+d}$, i.e. the fraction of energy $d(E_{\rm rad}/E)$ emitted per unit time and
per unit incoming beam flux $J\equiv dN/dA_{2+d} dt$ is given by:
$$
d\sigma_d =  \Omega_{1+d} b^{1+d}  db\,d(E_{\rm rad}/E)  =\frac{\Omega_{1+d}}{2(2\pi)^{3+d}m \gamma }
|\rho(k)|^2 b^{1+d} d b \, \omega^{2+d} d \omega \, d\Omega_{2+d}\, ,
$$
with the volume of the unit $(n-1)-$dimensional sphere being
$\Omega_{n-1}\equiv 2\pi^{n/2}/\Gamma(n/2)$.
Its integral over $b$ from $b_{\rm min}$ to infinity is the frequency and angular distribution of the
emitted energy fraction
per unit time and unit incoming beam flux, while
its total integral is the fraction of initial energy emitted per unit time and unit flux. Using
(\ref{sc12}) this is
\begin{align}
\label{sc12uuu}
\sigma_d= \frac{\Omega_{1+d}}{m \gamma}\int \limits_{b_{\rm min}}^{\infty}
dE_{\rm rad} \, b^{1+d}\, db
\simeq {\mathcal C}_d\, \frac{\Omega_{1+d}}{1+2d}\, r_d^{2+d}\, \gamma^{1+d}\,
\left(\frac{r_d} {b_{\rm min}}\right)^{1+2d}\,.
\end{align}

\subsubsection{The average number $ \langle N \rangle$ of emitted quanta}

A useful, though {\it not classical}, quantity to estimate is the number of emitted quanta.
Divide the right hand side of (\ref{pafinal}) by the quantum energy $\hbar\omega$ to obtain
$$
\label{N_aver_Mink}
\frac{d {\langle N \rangle}}{d\omega d\Omega} \sim
\frac{1}{2(2 \pi)^{3+d}} \frac{\omega^{1+d}}{\hbar} |\rho (k)|^2
\Big|_{k^M=\omega (1,\mathbf{n})},
$$
substitute (\ref{rho_u0}) and integrate, as above, over frequencies and angles to obtain an
estimate of the total number of emitted $\Phi-$quanta, not counting the ones from $m'$
with very low $\omega\to 0$ (see Figure \ref{primed_vs_unprimed})
\begin{align}\label{N}
{\langle N \rangle} \sim \frac{\Gamma\left(\frac{2+3d}{2} \right) \Gamma^2\! \left(1+d\right)
\Gamma^2\!\!\left(\frac{2+d}{2} \right)}{(3+d)\,2^{5-d} \pi^{4+3d/2} \Gamma(2+2d)\Gamma(2+d)}\,
\frac{m\,r_d}{\hbar}\,\gamma^d
\,\left(\frac{r_d}{b}\right)^{2+3d} \,.
\end{align}

\section{Scalar ultra-relativistic bremsstrahlung in
$\mathbf{{\mathcal M}_4 \times  {\mathcal T}^{\, \textit d}}$}
\label{Kaluza}

\subsection{The model - General formalism}

The above analysis is easily extended to study bremsstrahlung in the context of a scalar
multidimensional model with the extra dimensions compactified on a torus ${\mathcal T}^d$.
The model of interest describes two particles on the brane with masses $m$ and $m'$,
respectively, and two scalar fields. One, the "scalar graviton" $\Phi$, lives in the $D=4+d$ dimensional
bulk and is supposed to imitate gravity, while the other, $\phi$, lives
on the 3-brane and mimics the Standard Model forces. The action of the model is
\begin{equation}
\label{actionsc}
S_{\rm sc}= \frac{1}{2}\int d^4x \partial_{\mu}\phi\, \partial^{\mu}\phi\,
+\frac{1}{2}\int d^Dx \partial_{M}\Phi\, \partial^{M}\Phi
- \sum_{\rm particles} \int d\tau \sqrt{\zt^2}\left[m+e \phi(z)+f\Phi(z) \right]
\end{equation}
with $\mu=0, 1, 2, 3;\;M = 0,1, \ldots ,D-1$. $z^\mu(\tau)$ and $z'^\mu(\tau')$ are the trajectories of the
particles with masses $m$ and $m'$, respectively.
Notice that no $\Phi - \phi$ interaction is included, because it is not relevant in the leading
order computation that follows. The coupling constants $e, e'$ are
dimensionless, while $f, f'$ have length dimension $d/2$. With all radii of the
torus taken for simplicity equal and denoted by $L$, the kinetic term of the zero mode of $\Phi$ is
$(L^d/2) \int d^4x \partial_\mu \Phi \partial^\mu \Phi$. It is
brought to the standard normalization by the rescaling $\Phi \to L^{-d/2} \Phi$,
which converts the interaction term to $f_4 \int d\tau \sqrt{\zt^2} \Phi$ with the 4-dimensional coupling
$f_4$ being
\be
f_4=f L^{-d/2}\,.
\label{f4}
\ee
The particles, apart
from the radius $r_d$ defined in (\ref{rd}), are also characterized by the "electromagnetic"
classical radius, defined by
\be
R_0\equiv e^2 / m\,,
\ee
and similarly for the particle $m'$ \footnote{In the quantum theory, one may trade $f$ for a
"Planck" scale $M_*$ writing $f=(M_*/\hbar)^{-d/2}$. Then, one obtains
$f_4=(M_* L/\hbar)^{-d/2}$, the analog
of the ADD relation between "Newton's" constant, "Planck" scale and internal space volume.}.

This simplified model is rich enough to study all four cases presented in Table \ref{tableM4Td}, namely,
the energy emitted in the bulk or on the brane, when the scattering
of the two particles is dominated by either the $\Phi$ or the $\phi$ interactions.
The cases of $\phi-$interaction with $\phi-$radiation on the brane and $\Phi-$ interaction with
$\Phi-$radiation in the bulk are special cases of the
previous section. The formalism necessary for the study of the crossed situations i.e.
interaction via $\Phi(\phi)$ and $\phi(\Phi)-$radiation in the bulk, is the topic discussed next.

\vspace{0.5cm}

{\it Retarded propagator for massive modes and particle trajectories.}
Denote the space-time coordinates by $ x^M=\{x^\mu, y^i \}$, with $\mu=0,1,2,3;\;
i=1,\ldots,d$ and use bold letters to denote the $d-$vectors or the extra components of the
$D-$vectors.

The retarded Green's function of the d'Alembert equation
\begin{align}
\label{KKb0}
\Box_D  G_{D}(x,x',\y-\y')=  \delta^4(x-x')\delta^d(\y-\y')
\end{align}
is now expanded in Fourier series:
 \begin{equation}
 \label{KK1a22_mi1}
 G_{D}(x-x',\y-\y') =\frac1{(2\pi)^4 V_d}
 \int d^4 p \, \e^{-ip\cdot (x-x')} \sum_{\n \in
{\mathbb{Z}}^d} \frac{\e^{i\bbn \cdot (\y-\y')}}{p^2-\brmn^2+i \epsilon p^0},
 \end{equation}
where $\bbn \equiv  2\pi\n/L,\; \brmn \equiv \sqrt{\bbn^2}$, and $V_d=L^d$ is the volume
 of extra space. Correspondingly, the solution of
\begin{equation}
\label{Bobr}\Box_D \Phi(x,\y)=\rho_D(x,\y)
\end{equation}
 with the source localized on the brane \footnote{In this section we have to distinguish
the $D$-dimensional scalar source $\rho_D$ defined in the whole
$D=4+d$ space and the four-dimensional $\rho(x)$ localized on the
brane.}
\begin{equation}
\rho_D(x,\y )=\rho(x)\delta^d(\y)
\end{equation}
is
 \begin{align}
 \Phi(x,\y)=\frac{1}{(2\pi)^4 V_d}\int d^4 p \, e^{-ip\cdot (x-x')}\sum_{\n \in
\mathbb{Z}^d}\frac{\e^{i\bbn\cdot \y} }{{p}^2-\brmn^2+i \epsilon p^0 }\;
\rho(x')\;d^4x'.
 \end{align}
Its restriction to the 3-brane  can be rewritten using the
four-dimensional propagator
\begin{align}
\label{KKb2}
\Phi(x)\equiv\Phi(x,\y)\Big|_{\y=0}=\int G_{4}(x-x')\rho(x') d^4
x',\quad G_{4}(x-x')=G_{D}(x-x',\y-\y')\left.
\vphantom{\sqrt{f}}\right|_{\y=\y'=0},
\end{align}
whose Fourier transform is
\begin{align}
\label{KK1a22_mi1qq}
G_{4}(p)=\frac{1}{V_d}\sum_{\n \in \mathbb{Z}^d}
\frac{1}{{p}^2-\brmn^2+i \epsilon p^0}.
\end{align}
In  the four-dimensional language the massive scalar KK modes act
as independent massive fields universally interacting with scalar
charges through the total field
 \begin{equation}
 \Phi(x)=\sum_{\n \in \mathbb{Z}^d}\Phi_\n(x).
 \end{equation}
Thus, keeping four-dimensional conventions for the Fourier-transform, we
can present the fields generated by the particle $m$ to lowest
order as
\begin{align}
\label{ooo1}
  \Phi_\n(q)= - \frac{2\pi f}{V_d}
  e^{iq\cdot z(0)}\delta(q\cdot u)\frac{1}{q^2-\brmn^2},
\end{align}
and similarly for   $m'$. Taking into account only mutual particle
interactions one obtains the lowest order corrections to the
particles' trajectories:
\begin{align}
\label{sc6} \delta z^{\mu}(\tau)= -i \frac{ f f'}{m(2\pi)^{D-1}V_d}
\sum_{{\bf n} \in \mathbb{Z}^d}\int d^4 q
\frac{1}{q^2-\brmn^2}\frac{1}{(q\cdot u)^2}\, \e^{-iq\cdot
b}(e^{-i(q\cdot u)\tau}-1+i q\cdot u \tau) \delta(q\cdot u') \Pi^{\mu\nu} q_{\nu}.
\end{align}
It can be checked that the corresponding corrections in the
transverse directions
$$
\delta \mathbf{y}= -i \frac{  f f'}{m
(2\pi)^{D-1}V_d} \sum_{\n \in \mathbb{Z}^d}\int d^4 q
\frac{\bbn}{q^2-\brmn^2}\frac{1}{(q\cdot u)^2}\, \e^{-iq\cdot
b}e^{-i(q\cdot u)\tau} \delta(q\cdot u') = 0
$$
by parity and, as expected, the particles do not leave the brane.

\vspace{0.5cm}
{\it $\Phi-$radiation flux.}
Consider a $(D-1)$-dimensional space-like hypersurface in the
$D-$dimensional space-time and choose it for simplicity to be
orthogonal to the time axis. The total $D-$momentum of the field
$\Phi(x,\y)$ associated with it reads:
 \begin{equation}
 \label{Mom5}
 P^M=\int_{\mathbb{R}^3\times T^d}  T^{M0} d^3 x d\y.
 \end{equation}
The field   generated by a source localized on the brane is
  \begin{equation}
  \Phi(x,\y)=\frac{1}{(2\pi)^4 V_d}\int d^4 k \sum_{\n \in \mathbb{Z}^d}
 \frac{\rho(k)\e^{-ik\cdot x+i\bbn\cdot
 \y}}{k^2-\brmn^2+i\epsilon k^0},
  \end{equation}
where $\rho(k)$ is the four-dimensional Fourier-transform of the
source. Substituting this into (\ref{Mom5}) and integrating over the
three space and the  torus, one obtains for $M=i$:
 \begin{equation}
 \mathbf{P}(t)\sim\int dk^0 dk'^0 \sum_{\n \in \mathbb{Z}^d}\frac{\rho(k^0,\underline{\bf k})
 \rho^*(k'^0,\underline{{\bf k}})\e^{i(k^0-k'^0)t}k^0 \bbn}{[(k^0)^2-\underline{{\bf k}}^2-\brmn^2]
 [(k'^0)^2-\underline{{\bf k}}^2-\brmn^2]},
 \end{equation}
where $\underline{{\bf k}}$ is a wave 3-vector on the brane. The sum
vanishes by parity and reflects the expected fact that the
radiation is emitted in the bulk symmetrically with respect to the brane.

Similarly, the change of the tangential to the brane
components of the momentum of  $\Phi$ between two  space-like
hypersurfaces $\Sigma_{\pm\infty}$ of topology
$\mathbb{R}^3\times {\mathcal T}^d$ is again given by
(\ref{momecha}), or equivalently
 \begin{equation}P^\mu=\int d^4x\int_{V_d} d\y \nabla_N T^{\mu N},
 \end{equation}
from which
 \begin{equation}
 P^\mu=\int d^4x\int_{V_d} d\y \; (\rho_D+\rho_D')\; \partial^\mu \Phi\,,
 \end{equation}
 with the sources localized on the brane. Thus, the integral
 reduces to a four-dimensional one, and the retarded scalar field can be
computed using the four-dimensional Green's function. In particular, the energy emitted is
\begin{align}
\label{sc_pert10ma}
P^{0}   \equiv E_{\rm rad} =  \frac{1}{16 \pi^{3} V_d}\sum_{\n \in
\mathbb{Z}^d} \int\limits_{0}^{\infty}\!\! \omega^{2} d\omega
\int\limits_{S^{2}}\!\! d\Omega
    \left. \vphantom{\sqrt{d}} |\rho^{\tot}(k)|^2
    \right|_{k^0=\sqrt{\omega^2+\brmn^2}},
\end{align}
where $\omega=|\underline{\mathbf{k}}|$ and $\rho^{\tot}(k)$ is
the four-dimensional Fourier-transform of the source.

\subsection{Particles interact via $\Phi$ and emit $\phi-$radiation on the brane}

Consider first the case in which the colliding particles interact mainly via $\Phi$. Technically one may
take $e'=0$ in (\ref{actionsc}).
To compute the $\phi-$radiation emitted on the brane by particle $m$, one has to solve
the field equation for $\phi$ with the source term of order $ff'$. The radiation amplitude is the
Fourier-transform of the source $\rho(k)$ on the
mass-shell $k^2=0$ of $\phi$, while  the radiation energy loss will
be given by the $\bf n=0$ term in (\ref{sc_pert10ma}), or
equivalently, by (\ref{pafinal}) with $D=4$ and $\rho'=0$.

\subsubsection{The emitted energy}

The source $\rho(k)$ is of the form (\ref{rho3+3}), with an obvious change in the couplings
in front and summed over all KK modes. Using (\ref{I_mm_vec},\ref{I_mm_s}) relevant to
the contribution of the generic massive mode $\bf n$, one obtains
\begin{align}
\label{sc8} \rho(k)=\frac{ e f f' }{2\pi m   L^d \gamma^3 v^4 z^2}
\sum_{\n \in \mathbb{Z}^d}  \left(b z K_0(Z_{\brmn}) (\gamma
k\cdot u - k\cdot u' ) +i v \hat{K}_{1}(Z_{\brmn}) k\cdot b
\right),
\end{align}
with $Z_{\brmn}\equiv \sqrt{z^2+\brmn^2 b^2}$.

The sum over the massive modes depends on
the ratio $b/L$. For small $b/L$ a large number of modes will be
excited and for a summand that depends only on the magnitude of $\brmn$ one can write (see Appendix
III for an estimate of the error due to this approximation)
\begin{align}
\label{sum2int}
 \sum_{\n} f_\n \approx \int f_\n d\n=\frac{V_d}{(2\pi)^d}
 \Omega_{d-1}\int f_{\brmn} \brmn^{d-1}
 d\brmn.
\end{align}
Define the new variable $w\equiv 1+(\brmn b/z)^2$ and use formula
\cite{GR}  (for $a, \mu>0$)
\begin{align}
\label{int_modes} \int\limits_1^{\infty}\check{K}_{\nu}(a
\sqrt{w})(w-1)^{\mu-1}dw=2^{\mu}a^{-2\mu}
\Gamma(\mu)\check{K}_{\nu-\mu}(a)
\end{align}
to obtain
\begin{align}
\label{sum2int1} \sum_{\n \in \mathbb{Z}}\hat{K}_{\nu}(Z_{\brmn})
\approx \frac{(L/b)^d}{(2\pi )^{d/2} } \hat{K}_{\nu+d/2}(z),
\qquad \nu\geqslant 0,
\end{align}
where $\check{K}_{\nu}(y)\equiv \hat{K}_{-\nu}(y)$.
The resulting expression for the source coincides
with the $(4+d)$-dimensional (\ref{rho3}). According to the reasoning of Section II,
the leading contribution to the emitted energy in the ultrarelativistic case comes from
the first term inside the parenthesis of (\ref{sc8}) and gives
\begin{align}
\label{rtisc1} \rho^{\tot}(k)\simeq  \frac{e f f'}{(2\pi )^{d/2+1}
\, m v^3 \gamma^3 b^d } (\gamma^2 - \xi^{-1}) \hat{K}_{d/2}(z).
\end{align}
Thus, with an appropriate correspondence of the couplings, the source looks \emph{the same} as
for the massless field in the case of non-compactified space-time of
dimension $D=4+d$. The total emitted energy, however, will be different because
of the different phase space in the two cases.

\begin{figure}
\centerline{
\includegraphics
[angle=0,width=12cm]{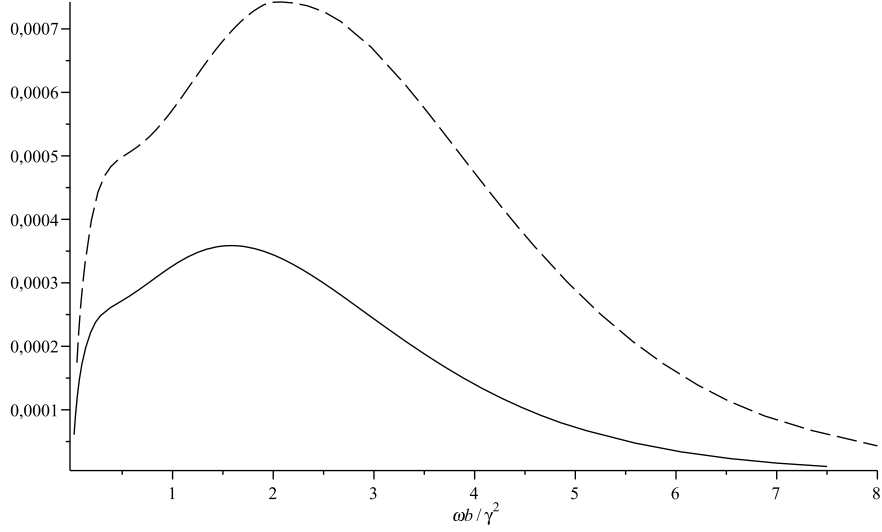}}\caption{Frequency distribution
$d\varepsilon_d/ dy$ (d=1, solid curve) and $\gamma d\varepsilon/dy$
(d=2, dashed) versus $y\equiv b\,\omega/\gamma^2$, both for
$\gamma=500$, $r_d/b=R_0/b=0.3$. \label{ADD1_omega_distr}}
\end{figure}

Substitution of (\ref{rtisc1}) into (\ref{sc_pert10ma}) and integration
over the frequencies leads to the angular distribution
$$
\frac{d E_{\rm rad}}{d
\Omega}=\frac{\Gamma\left(\frac{3+d}{2}\right)\Gamma\left(\frac{3+2d}{2}\right)}
{2^5 (2\pi)^{4+d}\, \Gamma\left(\frac{4+d}{2}\right)} \frac{ e^2 f^2
f'^2 }{ m^2\gamma^6 b^{3+2d}}\,\xi^{-5}\,(1-\gamma^2 \xi)^2,
$$
or equivalently (for $f=f'$ and $m=m'$) \be
\frac{d\varepsilon_{\Phi\phi}}{d \Omega}
=\frac{\Gamma\left(\frac{3+d}{2}\right)\Gamma\left(\frac{3+2d}{2}\right)}
{2^5 (2\pi)^{4+d}\, \Gamma\left(\frac{4+d}{2}\right)}
\frac{R_0}{b}\, \left(\frac{r_d}{b}\right)^{2+2d} \,
\gamma^{-7}\,\xi^{-5}\,(1-\gamma^2 \xi)^2. \label{dedOmega} \ee
Thus, up to the overall coefficient the angular distribution is the
same as in $4D$ shown in Figure \ref{theta_distr}. No sign of extra
dimensions in the angular profile of the radiation emitted on the
brane, in contrast to the frequency distribution shown in Figure
\ref{ADD1_omega_distr}.

Finally, use
$$
\int\limits_{0}^{2\pi} d \varphi  \int\limits_{0}^{\pi} \sin
\theta \, \xi^{-5}\,(1-\gamma^2 \xi)^2\,d\theta =
\frac{4 \pi}{3}\gamma^{8}+\mathcal{O}(\gamma^{6}),
$$
to integrate over the angles and obtain for the efficiency in the ultra-relativistic limit:
\begin{align}
\label{rtisc3a}
\varepsilon_{\Phi\phi} \simeq
{\mathcal C}_{\Phi\phi}\,\frac{R_0}{b}\left(\frac{r_d}{b}\right)^{2(1+d)} \gamma\,,
\end{align}
with
\be
{\mathcal C}_{\Phi\phi}\equiv \frac{\Gamma\left(\frac{3+d}{2}\right)\Gamma\left(\frac{3+2d}{2}\right)}{3
\cdot 2^4 (2\pi)^{3+d}\, \Gamma\left(\frac{4+d}{2}\right)}\,.
\ee
As a check, notice that for $d=0$ and the identification $e=f$ it coincides with
(\ref{sc12}).

\subsubsection{The cross-section $\sigma_{\Phi \phi}$}

For the scattering process, which takes place on the 3-brane, the total cross-section for ultra-relativistic
scattering is the integral of the differential energy
cross-section
\be
d\sigma_{\Phi \phi} =  2\pi b  \, db \, d\, \!E_{\rm rad} / m \gamma,
\label{4dsigma}
\ee
with $dE_{\rm rad}(\omega, \Omega, b)$
given above. Integration over the energy $\omega$ of the emitted scalar and
the angles $\Omega=(\theta,\varphi)$ on the brane leads to $E_{\rm rad}(b)$
and upon integration over $b$ to (for $f=f'$ and $m=m'$)
\be
\label{rti6}
\sigma_{\Phi \phi}
\simeq \frac{2\pi}{1+2d}\,{\mathcal C}_{\Phi\phi}\,r_d^2\,\frac{R_0}{b_{\rm min}}
\left(\frac{r_d}{b_{\rm min}}\right)^{2d}\, \gamma \,.
\ee
For $d=0$ it coincides with the corresponding expression derived in Section II.

\subsection{Interaction via $\phi$ on the brane, emission of $\Phi$ in the bulk}

Consider the action (\ref{actionsc}) with $f'=0$. Now the
particles interact only via the brane field $\phi$, but radiate
both on the brane ($\phi$ and massless mode of $\Phi$) and in the
bulk (massive modes $\Nv$ of $\Phi$). One wishes to estimate the
amount of radiation into the bulk.

\subsubsection{The energy radiated in the bulk}

Following steps analogous to the ones above, one starts with the source for the $N-$th mode
(eqn. (\ref{rho3}) with $D=4$)
\begin{equation}
\label{rho3b} \rho^{(\Nv)}(k)=-\frac{ e e' f }{2\pi \,L^d
\,m\gamma^3\, v^3 } \left[ \left( \frac{\gamma z'}{ z}
-\gamma^2\right) {K}_0(z)+i v
\sin\theta\sin\psi\hat{K}_{1}(z)/z\right],
 \end{equation}
with the arguments of the Macdonald functions in $\rho(k)$ and
$\rho'(k)$ being
 \begin{equation}\label{Elzin}
 z\equiv \frac{k \cdot u b}{\gamma v}=\frac{b}{v}\left(\sqrt{\omega^2+{\bar N}^2}-\omega v
 \cos\theta\right) \qquad {\rm and} \qquad  z'\equiv \frac{k\cdot u' b}{\gamma v}=\frac{b}{\gamma v}\sqrt{\omega^2+{\bar N}^2}\,,
 \end{equation}
depending on the mode vector $\Nv.$

Next, one has to substitute (\ref{rho3b}) into
(\ref{sc_pert10ma}), integrate over frequencies and angles and,
finally, sum over the KK tower.

As in the previous cases, the main contribution to the emitted energy is due to
the real part of the fast particle's source.
To compute it, introduce $\Xi\equiv k\cdot u/ (\gamma k\cdot u')$ and write
 \begin{equation}
 E_{\rm rad}= \frac{e^2 e'^2 f^2 \Omega_{d-1} }{2(2\pi)^{4+d}
m^2\gamma^6} \sum_{\Nv \in \mathbb{Z}^{d}} \int_0^{\pi}
 \sin\theta \,
d\theta \left(\Xi^{-1}-\gamma^2\right)^2 \int_0^\infty \omega^2
d\omega K_0^2 (z).
 \end{equation}
Isolate the contribution
of the massless mode and use (\ref{sum2int}) to convert to integration over masses.
Then change to polar coordinates $(w,\alpha)$ by
$\omega=w\cos\alpha,\; \bar N=w \sin\alpha$ with $0<\alpha <
\pi/2$ and defining $\Xi=1-v\cos\alpha \cos
\theta,$ integrate over $w$:
$$\int_0^\infty
dw w^{d+2}\,K_0^2(w b \Xi /v)=\frac{(v/b \Xi)^{d+3}2^d
}{\Gamma(d+3)} \Gamma^4\left(\frac{d+3}{2}\right).$$
Then  the leading in $\gamma$ contribution to the total emission in the bulk is given by
\begin{align}
 E_{\rm rad}= F \int\limits_0^{\pi/2} \cos^2 \alpha\, \sin^{d-1}\alpha
d\alpha \int\limits_{0}^{\pi} \frac{d\theta \sin\theta}{\Xi^{d+5}}
\left(1- \gamma^2 \Xi\right)^2 \;,\;\; {\rm with} \;\; F\equiv
\left(\frac{e e'f }{m \gamma^3}\right)^2\frac{
\Omega_{d-1}\,2^{d-1}\,
\Gamma^4\left(\frac{d+3}{2}\right)}{(2\pi)^{4+d}\, \Gamma(d+3)\,
b^{d+3}}.
\end{align}
The integral over $\theta$ can be converted into an integral over
$\Xi$ from $\xi_{\alpha} \equiv 1-v \cos\alpha$ to
$\bar{\xi}_{\alpha} \equiv 1+v\cos\alpha.$ The range of integration over $\alpha$ is such that
one can omit the terms with
$\bar{\xi}_{\alpha}$, and write
\begin{align}
\label{gmyr1}
E_{\rm rad} \simeq F \int\limits_0^{\pi/2}  \cos \alpha  \sin^{d-1}\alpha \,
d\alpha \left(\frac{\xi_{\alpha} ^{-(d+4)}}{d+4}-2 \gamma^2 \frac{
\xi_{\alpha}^{-(d+3)}}{d+3}+\gamma^4
\frac{\xi_{\alpha}^{-(d+2)}}{d+2}\right),
\end{align}
with all remaining integrals of the form $V_m^n$ (\ref{jj2q}) of the first kind ($2m>n+1$).
The difference in the range of integration is insignificant, because it leads to
${\mathcal O}(\gamma^{-2})$ contribution in the integrals, which is negligible in
the ultra-relativistic limit. Change the range of $\alpha$ integration to $[0,\pi]$ and use
(\ref{jj2q}) to evaluate (\ref{gmyr1}). To leading order in $\gamma$ the result is (for $e=e'$ and $m=m'$)
\begin{align}
\label{gmyr3} \varepsilon_{\phi\Phi} \simeq  {\mathcal
C}_{\phi\Phi}\, (ee'f_4)^2 \left(\frac{L}{b}\right)^d
\frac{\gamma^{d+1}}{m^3\, b^3} \simeq {\mathcal C}_{\phi\Phi}\,
\frac{R_0^2}{b^2}\,\left(\frac{r_d}{b}\right)^{d+1}\, \gamma^{d+1},
\end{align}
with \be {\mathcal
C}_{\phi\Phi}=\frac{\Gamma^3\left(\frac{d+3}{2}\right)}{2^5
\,\pi^{(d+7)/2}\, \Gamma(d+4)}. \ee

\subsubsection{The cross-section $\sigma_{\phi\Phi}$}

Upon multiplication of (\ref{gmyr3}) by $2\pi b\,db$ and integration over $b$
one obtains the total cross section
$$
\sigma_{\phi \Phi} \simeq \frac{2\pi}{1+d}\, {\mathcal C}_{\phi\Phi}\,
r_d^2\, \frac{R_0^2}{b_{\rm min}^2} \left(\frac{r_d}{b_{\rm min}}\right)^{-1+d} \gamma^{1+d}\,.
$$


\subsection{Interaction via $\Phi$ and $\Phi-$radiation in the bulk.}

Consider, finally, the case in which the particles interact via exchange of $\Phi$ and
emit $\Phi-$radiation in the bulk. Their couplings to $\Phi$ are
$f$ and $f'$, respectively. The source of radiation of the ${\bf N}-$th mode is then:
\begin{align}
\label{sc8full} \rho^{(\Nv)}(k)=-\frac{f^2 f' }{2\pi m\gamma^3 v^3
L^d}\sum_{\n} \left( \left( \frac{\gamma z'}{ z} -\gamma^2\right)
{K}_0(Z_{\brmn})+i v^2
\frac{\omega\sin\theta\sin\psi}{\sqrt{\omega^2+{\bar N}^2} z}
\hat{K}_{1}(Z_{\brmn}) \right),
\end{align}
with the argument of the Macdonald functions being
 \begin{equation}
Z_{\brmn}=b \left[\left(\sqrt{\omega^2+{\bar N}^2}-\omega v
 \cos\theta\right)^2 \!\!\! \left. \vphantom{\sqrt{f}}  \right/v^2    +\brmn^2 \right]^{1/2}
 \end{equation}
and the dimensionless products $k\cdot ub/\gamma v$ and $k\cdot u'
b/\gamma v$ given by (\ref{Elzin}).

 Assuming again a large
number of interaction modes, we replace the sum over modes by
integration according to (\ref{sum2int}):
\begin{align}
\label{hmyr0} \rho^{(\Nv)}(k)=-\frac{f^2 f' }{(2\pi)^{d/2+1}
m\gamma^3 v^3 b^d} \left( \left( \frac{\gamma z'}{ z}
-\gamma^2\right) \hat{K}_{d/2}(z)+i v^2
\frac{\omega\sin\theta\sin\psi}{\sqrt{\omega^2+{\bar N}^2} z}
\hat{K}_{1+d/2}(z) \right).
\end{align}
Thus, the  argument of the Macdonald function corresponds to the
massless interaction mode (\ref{rho3b}), so again the real part gives
the main contribution to the emitted energy, which in polar coordinates ($w,\alpha$)
in the $\omega -\bar N$  plane, becomes
\begin{align}
\label{hmyr0a} \rho^{(\Nv)}(k)=-\frac{f^2 f' }{(2\pi)^{d/2+1}
m\gamma^3 v^3 b^d} \left( \Xi^{-1}-\gamma^2\right)
\hat{K}_{d/2}(z),\quad \Xi\equiv 1-v\cos\alpha \cos \theta.
\end{align}
Substitute it in equation (\ref{sc_pert10ma}) and
integrate over $w$, using:
$$\int_0^\infty
dw w^{d+2}\hat{K}_{d/2}^2(w b \Xi)=\frac{\sqrt{ \pi}
\Gamma\left(\frac{3d+3}{2}\right) \Gamma
\left(\frac{2d+3}{2}\right) \Gamma\left(\frac{d+3}{2}\right)}{4
\Gamma(d+2) (b \Xi)^{d+3} }.$$
Perform the remaining integrations
over the angles $\varphi$, $\theta$ and $\alpha$ and use
similar approximations to obtain for the leading ultra-relativistic contribution to the
emitted energy ($f=f'$, $m=m'$)
\begin{align}
\label{hmyr2}
\varepsilon_{\Phi\Phi}
\simeq {\mathcal C}_d\, f_4^4 {f'}_4^2
\left(\frac{L}{b}\right)^{3d}
\frac{\gamma^{1+d}}{m^3 \,b^3} \simeq {\mathcal C}_d \left(\frac{r_d}{b}\right)^{3+3d} \gamma^{1+d}\,,
\end{align}
with ${\mathcal C}_d$ given in (\ref{CD}).
As expected, this expression is {\it identical} to the one obtained in the case of
$D-$dimensional Minkowski space (\ref{sc12}), even though the intermediate
formulae and numerical coefficients in the phase space integrals are different,
corresponding to different spatial topologies. This is a consequence of our
approximation to convert mode summation to integration.


The corresponding energy cross-section is
\be
\sigma_{\Phi \Phi} \simeq \frac{2\pi}{1+3d}\,{\mathcal C}_d\, f_4^4 {f'}_4^2 \left(\frac{L}{b_{\rm min}}\right)^{3d}
\frac{\gamma^{1+d}}{m^3\, b_{\rm min}}
\simeq \frac{2\pi}{1+3d}\,{\mathcal C}_d\, r_d^2
\left(\frac{r_d}{b_{\rm min}}\right)^{1+3d} \gamma^{1+d}\;.
\label{sigmaFF}
\ee

\section{Discussion - validity of the approximation - Prospects}

An unambiguous classical computation of bremsstrahlung radiation in ultra-relativistic
massive-particle collisions was presented in the context of a simplified scalar model in arbitrary
dimensions, of which some may be compact. Scalar fields were used to model both the graviton
in the bulk and the standard
model interactions on the brane and the main results for the radiation efficiency are
summarized in (\ref{varepsilonD}) and Table I.
A quick glance at these leads to the following remarks:
(a) In all cases there is enhancement of the efficiency by
factors of $\gamma$. (b) Radiation emitted on the 3-brane is enhanced by one power of $\gamma$, while
(c) each compact {\it large} extra dimension to which radiation can flow, contributes to the efficiency
an extra power of $\gamma$.

A few comments are in order concerning the validity of our approximations.
The leading order perturbative classical computation per se is a good approximation as long
as the interaction energy is much smaller than the total available energy, i.e.
\footnote{The discussion here concerns the case of
${\mathcal M}_D$. Similar analysis can be carried out in
${\mathcal M}_4\times {\mathcal T}_d$.} $U(b)\sim f^2/b^{1+d} \ll E\simeq m\gamma$. On the other hand,
the classical approach is relevant if one can justify (a) dealing with particle trajectories,
and (b) treating radiation classically. As has been discussed in non-relativistic quantum
mechanics \cite{landauIII} and applied to the
relativistic case as well \cite{giudice}, requirement (a) implies (a1) small angle scattering and (a2)
$|U'(b)|\, b^2 \gg \hbar$. Therefore, the impact parameter has to satisfy
\be
\frac{r_d}{\gamma^{1/1+d}} \ll b \ll \left(\frac{f^2}{\hbar}\right)^{1/d}\equiv b_c\, .
\label{condition1}
\ee
In addition, condition (b) of classicality of the radiation is written as
\be
\langle N \rangle \sim
\frac{m\, b}{\hbar\, \gamma}\,\left(\gamma
\,\frac{r_d^3}{b^3}\right)^{1+d} \sim \frac{m\,b}{\hbar \,\gamma} \, \varepsilon_d \gg 1\,.
\ee
Notice that for $\varepsilon_d\sim {\mathcal O}(1)$ the latter is satisfied for
$m\,b \gg \hbar\, \gamma$, which is equivalent to $\hbar\, \omega_{\rm max}
\sim \hbar \gamma^2 / b \ll m\gamma$,
which also has to be satisfied, since an emitted quantum cannot carry more than the
total available energy.
Finally, the deflection angle in the
ultrarelativistic case is given for small momentum transfers $|t|\ll s$ by \footnote{Note that: (a) The
cosine part of the exponential $exp(-iq\cdot u \tau)$ in
(\ref{z2}) does not contribute in $\delta\alpha$. (b) As explained
below, the points $\tau=0=\tau'$ on the trajectories are chosen so
that $u\cdot b = 0$.}
\begin{equation}
\delta\alpha=\frac{|\dot{z}\cdot \Delta|}{|\Delta|\gamma}\Bigg|^{\tau\to
\infty}_{\tau\to -\infty}=\lim_{\tau\to\infty}\frac{2i ff'}{m
(2\pi)^{D-1}b\gamma}\int d^D
q\frac{\e^{iq\cdot \Delta}\sin[(q\cdot u)\tau]\delta(q\cdot u')
(q\cdot \Delta)}{q^2(q\cdot u)},
\end{equation}
where $b\equiv |\mathbf{b}|$. Use $\displaystyle{\lim_{\tau\to\infty}\sin(q\cdot
u\tau)/(q\cdot u)=\pi\delta(q\cdot u)}$ and perform the integration
to obtain for $f=f'$
\begin{align}
\label{jj2rss1}
\delta\alpha=\frac{1}{\Omega_{1+d}} \left(\frac{r_d}{b}\right)^{1+d} \frac{1}{\gamma^2}\, .
\end{align}
The region of validity of this formula is $\delta\alpha\ll \sqrt{2/\gamma}\ll 1$ or equivalently
$b\gg r_d\, \gamma^{-3/2(1+d)}$,  and follows from the condition of small momentum transfer stated above.

For $\hbar \to 0$ the constraints are summarized to $b\gg  r_d\, \gamma^{-1/1+d}$. Taking into account
also the quantum constraints with $d\neq 0$
the impact parameter is restricted to $\hbar\gamma/m\ll b \ll b_c$, which
requires (for $\hbar=1$) $\gamma\ll f_4^{2/d} m L$.  In the special case of $d=0$ the region of validity of
our approximation is $b\gg \hbar\gamma/m$.

The results of the present paper are suggestive, but by no means conclusive about
gravitational bremsstrahlung itself. Gravity has many and significant differences from the toy scalar
model. (a) In gravity there is an
additional relevant scale $R_g= G_N \sqrt{s}$, which in principle enters the condition for the minimum
classically reliable value of the impact parameter.
(b) In addition, the fact that gravity couples to the energy-momentum is expected to lead
to extra enhancement of radiation in the transplanckian regime. Finally, (c) gravity is non-linear with
important effects due to these non-linearities. These qualitative features of classical transplanckian
gravitational bremsstrahlung have been verified and presented briefly in \cite{gkst3}, while a longer
detailed exposition is in preparation \cite{gkst4}.

\section*{Acknowledgements}
Work supported in part by the EU grants INTERREG IIIA
(Greece-Cyprus), MRTN-CT-2004-512914,
FP7-REGPOT-2008-1-CreteHEPCosmo-228644 and 08-02-01398-a of RFBR.
DG and PS are grateful to the Department of Physics of the University of Crete for its
hospitality in various stages of this work. TNT would like to thank the Theory Group of
CERN, where part of this work was done,
for its hospitality and also G. Altarelli and especially G. Veneziano for valuable discussions.

\section{Appendix I: Momentum integrals}

(a)  Start with the following invariant integral
\begin{align}
I \equiv m m' \int d^D q \frac{ \delta(q\cdot p')\,
\delta((k-q)\cdot p)\, \e^{-iq\cdot b}}{q^2}= \int d^D q \frac{
\delta(q\cdot u')\, \delta((k-q)\cdot u)\, \e^{-iq\cdot b}}{q^2}.
\end{align}
In the frame with $u'=(1,0,...,0)$ one may immediately integrate
over $dq^0 \delta(q^0)$ to obtain
\begin{align}
 I=-\int d^{D-1}
\mathbf{q} \frac{\delta(k\cdot u+\mathbf{q\cdot u})\,
\e^{i\mathbf{q\cdot b}}}{\mathbf{q}^2}.
\end{align}
Introduce $u^\alpha=\gamma(1,{\mathbf{v}})$, with
$\gamma=1/{\sqrt{1-v^2}}$ and decompose $\mathbf{q}$ and
$\mathbf{b}$ along $\mathbf{v}$ and perpendicular to it, writing
$\mathbf{q}=q_{\|}\mathbf{n}+\mathbf{q}_{\bot}$,
$\mathbf{b}=b_{\|}\mathbf{n}+\mathbf{b}_{\bot}$, where
$\mathbf{n}=\mathbf{u}/ \gamma v =\mathbf{v}/v$. It is
straightforward to check that
\begin{equation}
\label{covariantDelta} b_\|=\frac{1}{\gamma v}\left(\gamma u'\cdot
b-u\cdot b\right)\;\;{\rm and}\;\;  |\mathbf{b}_\bot| =
\left(-b^2-\frac{[(u'\cdot b) u- (u\cdot b) u']^2}{(u\cdot u')^2
v^2} \right)^{1/2}.
\end{equation}
Integrating over $dq_{\|}$ with $\delta(k\cdot u+q_{\|}\gamma v )$ we
obtain
\begin{align}
\label{Igeneral} I=-\frac{1}{\gamma v }\e^{-ik\cdot u
b_{\|}/\gamma v }\int d^{D-2} \mathbf{q}_{\bot} \frac{
\e^{i\mathbf{q_{\bot}\cdot b}_{\bot}}}
{\mathbf{q}_{\bot}^2+(k\cdot u)^2/\gamma^2 v^2 }
=-\frac{(2\pi)^{n+1}}{\gamma v b^{2n}}\e^{-ik\cdot u b_{\|}/\gamma
v} \hat{K}_n(z),
\end{align}
where $n\equiv D/2-2$, $z \equiv k\cdot u b/\gamma v$,
$\hat{K}_{\lambda}(w)\equiv w^{\lambda}K_{\lambda}(w)$ and
$b\equiv |\mathbf{b_\bot}|$ now. For $b_\|\sim
\mathbf{u}\cdot\mathbf{b}=0$ (the case of interest in the main
text) the above simplifies to
\begin{equation}
\label{Ispecial} I=-\frac{(2\pi)^{n+1}}{\gamma v b^{2n}} \; {\hat
K}_n(z).
\end{equation}
Correspondingly, the primed integral obtained by a
$u\leftrightarrow u'$ and $b_M \leftrightarrow -b_M$ exchange,
\begin{align}
I' \equiv \int d^D q \frac{ \delta(q\cdot u) \delta((k-q)\cdot
u')\, \e^{iq\cdot b}}{q^2}=-\frac{(2\pi)^{n+1}}{\gamma v b^{2n}}
 \hat{K}_{n}(z')
\end{align}
with $z'\equiv k\cdot u' b/\gamma v$.

\vspace{0.5cm}

(b)  Consider next the vectorial integral
 \begin{equation}
 I_M \equiv \int d^D q \;
\frac{\delta(q\cdot u') \delta((k-q)\cdot u)\, \e^{-iq\cdot
b}}{q^2}\; q_M.
\end{equation}
 It may be computed
with the help of
\begin{equation}
\label{Iageneral} I_M=i\frac{\partial I}{\partial b^M}=i
\left(\frac{\partial I}{\partial b_\|}\frac{\partial
b_\|}{\partial b^M} +\frac{\partial I}{\partial b}\frac{\partial
b}{\partial b^M} \right)
\end{equation}
 Its explicit fully covariant
form is obtained using (\ref{Igeneral}) and (\ref{covariantDelta})
\begin{equation}\label{dIdDelta||} \frac{\partial
I}{\partial b_\|}=i\frac{k\cdot u}{\gamma v}I\;,\;\; \frac{\partial
I}{\partial b} =-\frac{I}{b}\frac{\hat{K}_{n+1}(z)}{\hat{K}_n(z)},
\end{equation}
where use was made of the formulae
\begin{align}
&K'_n(z)=-K_{n-1}(z)-\frac{n}{z}K_n (z) =-K_{n+1}(z)+\frac{n}{z}K_n(z),\nn \\
& \hat{K}'_n (z)=-z\hat{K}_{n-1} (z),\qquad \check{K}'_n(z)=-z
\check{K}_{n+1} (z).
\end{align}
Using
 \begin{equation}
 \label{dDelta||dDeltamu}
\frac{\partial b_\|}{\partial b^M}=-\frac{\gamma u'_M -
u_M}{\gamma v}\;,\;\; \frac{\partial b}{\partial b^M}=-\frac{1}{b}
 \left(b_M+\frac{(u'\cdot b-u\cdot b\;  \gamma)u'_M
+(u\cdot b -u'\cdot b \; \gamma)u_M}{(u\cdot u')^2 v^2} \right),
\end{equation}
one is led to the final result
\begin{equation}
\label{Iaspecial} I_M =-\frac{(2\pi)^{n+1}}{\gamma v b^{2n+2}}
\left(\frac{b z}{\gamma v}(\gamma u'_M-u_M) \hat{K}_n(z)+ i
 b_M \hat{K}_{n+1}(z) \right).
\end{equation}

\vspace{0.5cm}

(c) The integral
\begin{align}
I^{(A)}\equiv \int d^D q \frac{ \delta(q\cdot u')\delta((k-q)\cdot
u)\, \e^{iq\cdot b}}{(q\cdot u)^A \; q^2}=\int d^D q \frac{
\delta(qu')\delta((k-q)u)\, \e^{iq\cdot b}}{(q_0u_0-q_{\|}\gamma
v)^A \;q^2}
\end{align}
after integration over $dq^0 \delta(q^0)$ gives the extra factor
$(-\mathbf{q\cdot u})^{-A}$, which by virtue of
$d(\mathbf{q\cdot u})\delta(ku+ \mathbf{q\cdot u})/\gamma v$ gives $1/(k\cdot u)^A$. Thus
\begin{align}
I^{(A)}(k)=\frac{I}{(k\cdot u)^A}.
\end{align}
Similarly, for the vectorial integrals
\begin{align}
I_M^{(A)} \equiv \int d^D q \frac{ \delta(q\cdot
u')\delta((k-q)\cdot u)\, \e^{iq\cdot b}}{(q\cdot u)^A \;
q^2}q_M=\frac{I_M}{(k\cdot u)^A},
\end{align}

Similar relations hold for the primed integrals. The denominators are $(k\cdot u')^A$.

\vspace{0.5cm}

(d) In the case of ${\mathcal M}_4\times {\mathcal T}^d$ space-time, one is led to the invariant
integral
\begin{align}
I_{\brmn}= \int d^4 q \frac{ \delta(q\cdot u')\delta((k-q)\cdot
u)\,\e^{iq\cdot b}}{q^2-\brmn^2},
\end{align}
where $\brmn=\sqrt{\bbn^2}$ and $\bbn=2\pi \n/L,\,\n\in \mathbb{Z}^d$.
It is computed in the same way. After the
$q^0$-integration, the denominator becomes
 $\mathbf{q}^2+\brmn^2+(k\cdot u/\gamma v)^2,$ and leads to
\begin{align}
\label{I_mm}
 I_{\brmn}=-\frac{2\pi}{\gamma v}
{K}_{0}(Z_{\brmn}),
\end{align}
with  $Z_{\brmn}\equiv \sqrt{z^2+\brmn^2 b^2}$ and  $z=k\cdot
u\;b/(\gamma v)$. If $\mm$ represents the single one-dimensional KK
mass $m_l$, we will refer to $Z_l \equiv \sqrt{z^2+m_l^2 b^2}$ and
to the integral $I_l$ respectively.

\vspace{0.5cm}

(e)  The vectorial integral can be computed by differentiation and
the result is
\begin{align}
\label{I_mm_vec} I^{M}_{\brmn}\equiv \int d^4 q \frac{
\delta(q\cdot u')\,\delta((k-q)\cdot u)\,\e^{iq\cdot
b}}{q^2-\brmn^2}\; q^{M} =-\frac{2\pi  }{\gamma v b^{2}}\left(b z
{K}_{0}(Z_{\brmn})\frac{\gamma u'^{M}-u^{M}}{\gamma v}+i \hat
{K}_{1}(Z_{\brmn}) b^{M}\right).
\end{align}
Analogously, if the integrals have the factor in the denominator $(q\cdot u)^s$
of the integrand, give the additional factor $(k\cdot u)^{-s}$
by virtue of the delta-function $\delta((k-q)\cdot u)$, namely one obtains
\begin{align}
\label{I_mm_s} I^{(s)}_{\brmn}=I_{\brmn} \; (k\cdot u)^{-s} \qquad
I^{(s)M}_{\brmn}=I'^{M}_{\brmn}\; (k\cdot u)^{-s}.
\end{align}
Analogous relations hold for the primed integrals.


\section{Appendix II}
\subsection{Angular integrals}
(a) In the main text the following angular integrals
$\theta$ were needed for integer $m$ and $n$
\begin{align}\label{jj0}
V_{m}^n\equiv \int\limits_0^{\pi}\frac{\sin^n
\theta}{\xi^m}d\theta,\qquad \xi=1-v \cos{\theta}.
\end{align}
Making use of the formula \cite{Proudn}, valid for any real
$a>|b|$, and $ \mathrm{Re} \, {\nu}>0$:
\begin{align}\label{jj1pru}
 & \int\limits_0^\pi\frac{\sin^{2\nu-1}{\theta}}{(a+b\cos{\theta})^{\mu}}d \theta=
\left(- \frac{2}{b}\right)^{\nu-1/2}\!\!\!\!\sqrt{\pi}
\Gamma{(\nu)}
 (a^2-b^2)^{(\nu - \mu
 )/2-1/4}P^{1/2-\nu}_{\nu-\mu-1/2}\left(\frac{a}{\sqrt{a^2-b^2}}\right),
\end{align}
we express the result in terms of the associated Legendre function
$P^\mu_\nu(z)$. In our case  $a=1, b=-v,$ so:
\begin{align}\label{jj1s}
 & V_{\mu}^{2\nu-1}=
\left( \frac{2}{v}\right)^{\nu-1/2}\!\!\!\!\sqrt{\pi}
\Gamma{(\nu)}
 \gamma^{-\nu+ \mu
 +1/2}P^{1/2-\nu}_{\nu-\mu-1/2}\left(\gamma\right)=
 \left( \frac{2}{v}\right)^{n/2} \!\!\!\! \sqrt{\pi} \Gamma
 \left(\frac{n+1}{2}\right)
 \gamma^{-n/2+ \mu}P^{-n/2}_{n/2-\mu} (\gamma ).
\end{align}
For $\gamma \gg 1,$ one can use the asymptotic formula  \cite{GR}:
\begin{align}
\label{jj2}
P^{\lambda}_{\rho}(z)=\left(\frac{2^{\rho} \Gamma(\rho+1/2)}
{\sqrt{\pi}\Gamma(\rho -\lambda +1)} z^{\rho} +\frac{\Gamma(-\rho
- 1/2)} {2^{\rho+1}\sqrt{\pi}\Gamma(-\rho -\lambda)}z^{-\rho-1}\right) (1+O(1/z^2)).
\end{align}
For $2m>n+1$ one finds to leading order
\begin{align}\label{jj2q}
V_{m}^{n}= & \frac{2^{m-1}}{v^{n/2}\Gamma(m)}    \Gamma
 \left(\frac{n+1}{2}\right)
 \Gamma \left(m-\frac{n+1}{2}\right)
\gamma^{2m -n-1},
\end{align}
while for $2m<n+1 $
\begin{align}\label{jj3}
V_{m}^{n}= \frac{2^{n-m}}{v^{n/2}} \frac{\Gamma\left(\frac{n+1}{2}\right)
\Gamma\left(\frac{n+1}{2}-m\right)} { \Gamma(n -m+1)}.
\end{align}
In the case
  $2m=n+1$ an expansion of the integral is logarithmic.

\vspace{0.3cm}

(b) Another integral used in the main text is:
\begin{align}\label{intSN}
\int_{S^{D-3}} |\sin \psi|^N d \Omega = \frac{2 \pi^{ (D-3)/2
}\Gamma\left(\frac{{N+1}}{2}\right)}
{\Gamma\left(\frac{D-2+N}{2}\right)} .
\end{align}

\subsection{Integrals of products of Macdonald functions }
\label{Macsfreqs}
Computation of integrals over the frequency or the impact
parameter involving products of two Macdonald functions of the same argument
is performed using the formula  \cite{Proudn}:
\begin{align}\label{intfreq}
 \int\limits_0^{\infty}K_{\mu}(cz)K_{\nu}(cz)z^{\alpha-1}dz=
\frac{2^{\alpha-3}\Gamma \left(\frac{\alpha+\mu+\nu}{2} \right)
\Gamma \left(\frac{\alpha+\mu-\nu}{2} \right)\Gamma
\left(\frac{\alpha-\mu+\nu}{2} \right)\Gamma
\left(\frac{\alpha-\mu-\nu}{2} \right)}{c^{\alpha}
\Gamma(\alpha)}.
\end{align}
Actually, only such integrals are needed in this work.
More general integrals involving functions of different arguments
arise from interference terms. They can be computed using the formula
\begin{align}
\label{theormult}
K_{\nu}(\lambda z)=\lambda^{\nu}
\sum_{k=0}^{\infty}\frac{1}{k!}\frac{z^k}{2^k}(1-\lambda^2)^k
 K_{\nu + k}(z)\;,  \qquad |1-\lambda^2|<1.
\end{align}
The typical integral
\begin{align}\label{jjj1yyy}
\Theta=  \int\limits_0^{\infty}
K_{\nu}(z)K_{\nu'}(z')z^{m}z'^{m'}\omega^{D-2}d\omega,
\end{align}
where $z=z'\gamma\xi$, can be cast for $\gamma\xi<1$ (i.e. $0
\leqslant   \theta \lesssim  \sqrt{2/\gamma}$) into the form
\begin{align}\label{jjj2yyy}
 (\gamma\xi)^{m} \left( \frac{\gamma v }{\rho}\right)^{D-1}\int\limits_0^{\infty}
K_{\nu}(\gamma\xi z')K_{\nu'}(z')z'^{m+m'}z'^{D-2}dz',
\end{align}
and for $\gamma\xi>1$ ($\sqrt{2/\gamma} \lesssim \theta\leqslant
\pi$) to
\begin{align}\label{jjj3yyy}
 \frac{1}{(\gamma\xi)^{m'} } \left( \frac{\gamma v }{\rho}\right)^{D-1}\int\limits_0^{\infty}
K_{\nu}( z)K_{\nu'}(z/\gamma\xi)z^{m+m'}z^{D-2}dz.
\end{align}
One is interested in an estimate of (\ref{jjj1yyy}) in
powers of $\gamma\gg 1$. Use (\ref{theormult}) to write
\begin{align}\label{jjj4}
 \Theta \! = \!\left\{%
\begin{array}{ll}
  \displaystyle \!\!\!  (\gamma\xi)^{m+\nu} \left( \frac{\gamma v
}{\rho}\right)^{D-1}\sum_{k=0}^{\infty}\frac{1}{2^k k!}(1-\gamma^2
\xi^2)^k
 \int\limits_0^{\infty}z'^k K_{\nu + k}(z')  K_{\nu'}(z')z'^{m+m'}z'^{D-2}dz', & \hbox{$\gamma\xi<1$;} \\
   \displaystyle \!\! \!   \frac{1}{(\gamma\xi)^{m'+\nu'} } \left( \frac{v }{\xi
\rho}\right)^{D-1}\sum_{k=0}^{\infty}\frac{1}{2^k
k!}\left(1-\frac{1}{\gamma^2 \xi^2}\right)^k
 \int\limits_0^{\infty}z^k  K_{\nu}(z) K_{\nu' + k}(z) z^{m+m'}z^{D-2}dz \!\!, \; \;  & \hbox{$ \gamma\xi>1$.} \\
\end{array}%
\right.
\end{align}
Denote by $M\equiv m+m'+D-1$ and integrate over $z$ with the help of (\ref{intfreq}). The integral in
(\ref{jjj4},a) is equal to
\begin{align}\label{jjj4ff}
\textstyle \frac{2^{M-3+k}}{\Gamma(M+k)}\Gamma
\left(\frac{M+2k+\nu+\nu'}{2} \right) \Gamma
\left(\frac{M+2k+\nu-\nu'}{2} \right) \Gamma
\left(\frac{M+\nu'-\nu}{2} \right)\Gamma
\left(\frac{M-\nu'-\nu}{2} \right),
\end{align}
while the one in (\ref{jjj4},b) is related to the above by the exchange $\nu
\leftrightarrow \nu'$.

Given that for fixed $\lambda>0$ the function
$\Gamma(x)\Gamma(\lambda-x)$ is decreasing for $x$ in $[0, \lambda/2]$
and increasing in $[\lambda/2, \lambda]$, one obtains the
inequality
\begin{align}\label{jjj4gg}
\textstyle \Gamma \left(\frac{M+2k+\nu+\nu'}{2} \right) \Gamma
\left(\frac{M+2k+\nu-\nu'}{2} \right) < \Gamma(M+k)\Gamma(\nu+k),
\end{align}
which leads to the estimate
\begin{align*}\Theta < \left\{%
\begin{array}{ll}
   \!\!\!  2^{M-3} \Gamma \left(\frac{M+\nu'-\nu}{2} \right)\Gamma
\left(\frac{M-\nu'-\nu}{2} \right)(\gamma\xi)^{m+\nu} \left(
\gamma v/\rho\right)^{\scriptscriptstyle
D-1}\sum\limits_{\scriptscriptstyle k=0}^{\scriptscriptstyle
\infty} (1-\gamma^2 \xi^2)^k \Gamma(\nu+k)/k!, & \hbox{$\gamma\xi<1$;} \\
  \!\!\!  2^{M-3}  \Gamma \left(\frac{M+\nu-\nu'}{2}
\right)\Gamma \left(\frac{M-\nu'-\nu}{2} \right)
(\gamma\xi)^{-(m'+\nu')} \left( v /\xi \rho
\right)^{\scriptscriptstyle D-1} \sum\limits_{\scriptscriptstyle
k=0}^{\scriptscriptstyle \infty}
 \left(1-\frac{1}{\gamma^2 \xi^2}\right)^k \!\! \Gamma(\nu'+k)/k!, & \hbox{$ \gamma\xi>1$.} \\
\end{array}%
\right.\end{align*}
Both series are summable because
\begin{align}\label{jjj4st}
\sum_{k=0}^{\infty}\frac{1}{ k!}(1-a^2)^k \Gamma(n+k)=\frac{
\Gamma(n)}{a^{2n}},\qquad |a|<1,
 \end{align}
and give the following upper bounds, which are enough for our purposes:
\begin{align}
\label{jjj6}
 \!\!\! \Theta \! < \!\left\{%
\begin{array}{ll}
  \!\!\!  2^{M-3} \Gamma \! \left(\frac{M+\nu'-\nu}{2}
  \right)\Gamma \!
\left(\frac{M-\nu'-\nu}{2} \right)\Gamma(\nu) \left( \frac{\gamma
v }{\rho}\right)^{D-1}\!\!\!\!(\gamma\xi)^{m-\nu}\!\!, & \hbox{$\gamma\xi<1$;} \\
    \!\! \!   2^{M-3}  \Gamma \!\left(\frac{M+\nu-\nu'}{2}
\right)\Gamma \! \left(\frac{M-\nu'-\nu}{2} \right)\Gamma(\nu')
 \left(\frac{v }{\xi
\rho}\right)^{D-1} \!\!\!\!(\gamma\xi)^{\nu'-m'}\!\!, \; \;\; \; & \hbox{$ \gamma\xi>1$.} \\
\end{array}%
\right.\end{align} For $\nu=\nu'$ the estimate is continuous at
$\theta=\arccos((\gamma-1)/\gamma v).$

\section{Appendix III}
\subsection{Effective number of interacting $\Phi-$KK modes}
In this subsection we would like to count the effective number of contributing
interaction KK modes
of the bulk field $\Phi$, studied in Subsection III.B,  and estimate the error due to the replacement
of the summation over those by integration.

Start with
the case $D=5$, i.e. with one-dimensional mass sequence.
It was argued that the dominant contribution to the energy loss comes from
the first term in the parenthesis of (\ref{sc8}) and requires the evaluation of the two-parameter
integrals
\begin{align}
\label{jlob2a}
G(\alpha,\beta)=\int\limits_{0}^{\infty} \! K_{0}\!
\left(\!\!\sqrt{z^2+\alpha^2}\right) K_{0}\!
\left(\!\!\sqrt{z^2+\beta^2}\right)  z^{2} dz,
\end{align}
for $\alpha \geqslant 0, \beta \geqslant
0$ and $\alpha+\beta>0$. Using the integral representation \cite{GR}
\begin{align}
\label{i2}
K_{\nu}(x)K_{\nu}(y)=\frac{1}{2}\int\limits_{0}^{\infty}\exp\left[-\frac{t}{2}-\frac{x^2+y^2}{2t}\right]
K_{\nu}\left(\frac{xy}{t}\right)\frac{dt}{t},
\end{align}
we obtain the following approximation $\tilde{G}(\alpha,\beta)$ for
$G$:
\begin{align}
\label{jl4}
\tilde{G}(\alpha,\beta)\simeq \frac{\pi^2}{32}
e^{-(\alpha+\beta)}\left(1+\frac{2^7  \alpha^2 \beta^2}{\pi
(\alpha+\beta)^3} \right)^{1/2}.
\end{align}
In view of the exponential fall-off of $\tilde G$ it is natural
to separate the modes to {\it light} for $\alpha+\beta <2$ and {\it heavy} for $\alpha+\beta>2$. The choice
$\alpha+\beta =2$ for the boundary is justified by just looking at the
numerical plots. For the heavy modes one
can neglect the unity in the parenthesis of (\ref{jl4}) and write
\begin{align}
\label{jl5}
\tilde{G}(\alpha,\beta) \approx \frac{\sqrt{2} }{4}\pi^{3/2}
e^{-(\alpha+\beta)}\frac{\alpha  \beta }{(\alpha+\beta)^{3/2}}\;,
\qquad \alpha+\beta>2
\end{align}
Correspondingly, for the light modes one may take approximately
\begin{align}\label{jl6}
\tilde{G}(\alpha,\beta) \approx \frac{\pi^2}{32}
e^{-(\alpha+\beta)}, \qquad \alpha+\beta<2.
\end{align}
In the cases of interest here $\alpha=2\pi b\, l/L, \beta=2\pi b\,
l'/L$ with integer $l$ and $l'$. Thus, denoting by $K\equiv l+l',$ one sees that the integer part
\begin{align}
\label{intmodes}
K_c =[L/(\pi b)]
\end{align}
defines the boundary between light ($K<K_c$) and heavy ($K>K_c$) modes. If
$L<\pi b$, there are no light modes in this classification. In the
sequel only the case $L>\pi b$ ($K_c > 1$) will be discussed.

To estimate separately the contributions of the light (heavy)
interaction modes to the emitted energy, substitute the first term
of (\ref{sc8}) into (\ref{pafinal}) and integrate over angles,
using (\ref{jl6}) and (\ref{jl5}), respectively. For the light
modes one obtains (up to coefficients)
\begin{align}
\label{jl7}
\sum G_{\rm light}\simeq \frac{\pi^2}{32}\sum_{K=1}^{K_c} e^{-2\pi K
b/L}(K+1), \qquad K_c \geqslant 1.
\end{align}
It is easy to sum the geometric series for any $K_c$. In particular, for $K_c \gg 1$
the highest power in $K_c$ gives
\begin{align}
\label{jl8}
\sum G_{\rm light}\simeq \frac{\Lambda_{\rm light} \pi^2}{32} K_c^2,\qquad
K_c \gg 1,
\end{align}
with $ \Lambda_{\rm light} \equiv 1/4-3/(4\e^{2}) \thickapprox 0.148
\thickapprox 1/7.$
Similarly, the heavy modes give
\begin{align}
\label{jl7t} \sum G_{\rm heavy}\simeq \frac{\pi^{3/2} }{2 K_c^{1/2}}
e^{-(\alpha+\beta)}\!\!\!\! \sum_{K=K_c+1}^{\infty}\frac{
e^{-2K/K_c}}{K^{3/2}} \sum_{l=0}^{K}l(K-l)\simeq  \frac{\pi^{3/2}
}{12 K_c^{1/2}}\!\!   \sum_{K=K_c+1}^{\infty} \e^{-2K/K_c} K^{3/2}.
\end{align}
Replacing the sum over $K$ by integration from $K_c$ to $\infty$,
one obtains for large $K_c$:
\begin{align}
\label{jl7u}
\sum G_{\rm heavy}\simeq \frac{\Lambda_{\rm heavy} \pi^{3/2}}{3 \cdot
2^{9/2} } K_c^2, \qquad K_c \gg 1,
\end{align}
 with $\Lambda_{\rm heavy}=7/(\sqrt{2} \e^{2}) +3
\sqrt{\pi}\left[1-\Phi\left(\sqrt{2}\right) \right]/4\approx 0.73
\thickapprox 5/7 $, where $\Phi$ is the Laplace's error function. In
 the case  $L<\pi
b$ (no light modes) one should integrate from 0 to  $\infty$ and
substitute $K_c \to L/(\pi b)$ in (\ref{jl7u}).

Using (\ref{jl8}) and (\ref{jl7u}) one obtains the following ratio
for $d=1$:
\begin{align}
\label{jl7v} \frac{\Delta E_{\rm light}}{\Delta E_{\rm
heavy}}\approx \frac{3 \pi^{1/2}   \Lambda_{\rm light}   }{  2^{1/2}
\Lambda_{\rm heavy} }=0.76 \approx 3/4.
\end{align}
Thus, despite the fact that each heavy mode is exponentially suppressed
their total contribution is comparable to the one of the light modes.

The above generalizes to arbitrary $d>1$.
One has to integrate over the domain $W$ defined as
$M\equiv \brmn+\brmn'< 2 \pi K_c /L \equiv M_c$ which represents the
multidimensional bispherical octahedron (the coefficient $\pi^2/32$
is  omitted):
\begin{align}
\label{eeff5} \sum G^{(d)}_{\rm light}\simeq
\frac{(V_d\Omega_{d-1})^2}{(2\pi)^{2d}} \int\limits_{W}\e^{-2M/M_c}
(\brmn\, \brmn')^{d-1} d\brmn \, d\brmn' \simeq
\frac{\Omega_{d-1}^2}{2^{2d}} \mathrm{B}(d,d)\gamma(2d,2)\,K_c^{2d},
\end{align}
with $\gamma(x,y)$ is the lower incomplete gamma-function,
$\mathrm{B}(x,y)$ is the Euler beta-function. The corresponding heavy
mode contribution reads:
\begin{align}
\label{eeff6}
\sum G^{(d)}_{\rm heavy}
\simeq \frac{2^{7/2}(V_d\Omega_{d-1})^2 }{\pi^{1/2}(2\pi)^{2d}}
\int\limits_{\overline{W}}\frac{\e^{-2M/M_c}}{M^{3/2}} (\brmn \,
\brmn')^{d} d\brmn \, d\brmn'
\simeq \frac{\Omega_{d-1}^2\, \mathrm{B}(d+1,d+1)\, \Gamma(2d+1/2,2) }{2^{2d-7/2}\pi^{1/2}}
\,K_c^{2d}\,,
\end{align}
with $\Gamma(x,y)$ the upper incomplete gamma-function. Thus, the
ratio is
\begin{align}
\label{eeff7}
\frac{\sum G^{(d)}_{\rm light}}{\sum G^{(d)}_{\rm heavy}}
\simeq \frac{\pi^{1/2}
\mathrm{B}(d,d)\gamma(2d,2)}{2^{7/2}
\mathrm{B}(d+1,d+1)\Gamma(2d+1/2,2)}\, ,
\end{align}
a decreasing function of $d$.

Add (\ref{eeff5}) with (\ref{eeff6}) for the emitted energy in $D-$dimensions and compare to
the energy loss $\Delta E_4$ in the purely four-dimensional case to obtain the estimate
\begin{align}
\label{eeff8}
\frac{\Delta E_{4+d}}{\Delta E_{4}}\sim K_c^{2d}.
\end{align}
This is in line with Table I and explains the origin of the enhancement
in the entries of the first row (bulk interaction dominance), as compared to
the ones in the second row (the case of brane interaction dominance).

\subsection{Effective number of emission modes and the angular distribution}

Let us compare the role of KK emission modes, studied in
Subsection III.C, with that of KK interaction modes analysed in
the previous above. Passing to integration over the emission
modes, we have found that for ultrarelativistic velocities  the
angular-frequency distribution is the same as in the
non-compactified case: the typical frequency being $ \langle
\omega \rangle = \gamma^2/b$ and the emission angles within the
cone $\theta < \hat{\theta} = \arccos v$. From equation
(\ref{gmyr1}) one can see that only a finite number of emission KK
modes contributes to the total energy loss. Thus, unlike the case
of interaction modes, here one can determine the effective number
of emission modes from the energy loss right from the beginning.
Indeed, from $v< \cos \alpha <1$ and $\sin \alpha <\gamma^{-1}$,
one concludes that
$$\brmn < w \gamma^{-1}= 2\gamma /b, $$
so that the effective number of emission modes is
\begin{align}
\label{emmodes}
N_{\rm eff} = \frac{\gamma L}{\pi b}.
\end{align}
This is $\gamma$ times greater than the effective number of
interaction modes $N_{\rm eff} = \gamma K_c.$ Notice that the massive
arguments (with equal masses $\brmn$ of the scalar quantum) of the
Macdonald functions in the \emph{rest frame} of the fast particle,
do not depend on the emission angle:
\begin{align}
\label{susu0} Z_{\brmn}=\frac{b}{v}\sqrt{\omega'^2+\brmn^2
v^2},\qquad z^{(\brmn)}=\frac{b}{\gamma v}\sqrt{\omega'^2+\brmn^2},
\end{align}
with $\omega'=|\mathbf{k}|$ in this frame. From (\ref{susu0}) it is
clear that the effective number of modes (demanding the argument to
be not greater than unity) of $z^{(\mm)}$ is $\gamma v$ times larger
than the corresponding one of $Z_\mm.$ The overall effect  of
massive emission modes is found passing to integration assuming that
$N_{\rm eff}$ is large.

The angular distribution for  a specific KK-mode is more
complicated since $z^{(\bar N)}$ depends on $\omega, \bar N$ and
the angle $\theta.$ For light emission modes one has $\cos \alpha
> v$. The form of $\Xi=1-v \cos \alpha \cos \theta$ suggests the
introduction of the "effective velocity" $$v_\eff^{(\bar N)} =v
\cos \alpha,$$ which differs from unity by
$\mathcal{O}(\gamma^{-2})$. Thus, the angular and spectral
properties can be derived from the  effective quantities
$v_\eff^{(\bar N)} $ and $\displaystyle \gamma_\eff^{(\bar N)}
 \equiv \left[1- (v_\eff^{(\bar N)})^2 \right]^{-1/2}.$ For the light modes one can expand
\begin{align}
\label{sc_pert13ma}
z^{(\bar N)}=\frac{\omega\, b}{v}\left(\xi+\frac{{\bar N}^2}{2\omega^2}+\pp
\right)\,.
\end{align}
Thus, the correction is of the order of $\mathcal{O}(\gamma^{-2})$
and one can substitute the frequency by its average value:
\begin{align}
\label{sc_pert14ma} &\xi_{\eff}^{(\bar N)}=\xi+\frac{\bar N^2
b^2}{2\, \gamma^4}, \qquad\quad \cos \theta_{\eff}^{(\bar N)}=\cos
\theta-\frac{\bar N^2 b^2}{2\, v\, \gamma^4}, \nn
\\  & v_{\eff}^{(\bar N)}=v-\frac{\bar N^2 b^2}{2\,v\, \gamma^4}, \qquad\quad
\gamma_{\eff}^{(\bar N)}=\gamma\left(1-\frac{\bar N^2 b^2}{\gamma^2}
\right).
\end{align}
This leads to the effective replacement of the emission angle
$$\theta_{\eff}^{(\bar N)}=\sqrt{\theta^2+ \bar N^2\, b^2/v\, \gamma^4}.$$
 Therefore  the emission of a given  $\NNv$-mode is concentrated within the
 cone
\begin{align}
\label{sc_pert16ma}
\hat{\theta}_{\eff}^{(\bar N)}=\sqrt{\hat{\theta}^2+\bar N^2 \, b^2/v\,
\gamma^4}=\arccos v_{\eff}^{(\bar N)},
\end{align}
and the small-angle approximation is valid for all light modes.

The radiation flux of a single KK-mode $\Nv$   under
\emph{massless} interaction mode can be obtained substituting the
effective Lorentz factor $\gamma_{\eff}^{(\bar N)}$ into Eqn. (\ref{sc12}) (with $D=4$).
Note that although it is  two powers
of gamma smaller than the general result, the exact correction can
not be calculated by such analysis, because of the presence of  the
interference term with the product of two Macdonald functions with
different arguments. The exact results may be obtained in the
center-mass frame, where the interference term vanishes.

Total radiation flux can be estimated simply as the product of the
four-dimensional result by the total number massive emission modes:
$$E_{4+d}\simeq E_4 N_{\rm eff}^{\;\;\;d}\; .$$
Again, this explains the origin of the relative enhancement in the entries of the first column
(corresponding to bulk emission) of Table I, compared to the ones of the second
(radiation of the brane field).


\begin {thebibliography}{20}
\bibitem{iliopoulos} J. Iliopoulos,  {\it Following the Path of Charm: New Physics at the LHC.}
arXiv:0805.4768 [hep-ph]

\bibitem{ADD}
N.~Arkani-Hamed, S.~Dimopoulos and G.~R.~Dvali,
Phys.\ Lett.\ B {\bf 429}, 263 (1998) [arXiv:hep-ph/9803315];
Phys.\ Rev.\ D {\bf 59}, 086004 (1999) [arXiv:hep-ph/9807344];\\
I.~Antoniadis, N.~Arkani-Hamed, S.~Dimopoulos and G.~R.~Dvali,
Phys.\ Lett.\ B {\bf 436}, 257 (1998) [arXiv:hep-ph/9804398].

\bibitem{ablt}
  I.~Antoniadis, C.~Bachas, D.~C.~Lewellen and T.~N.~Tomaras,
  Phys.\ Lett.\ B {\bf 207}  441 (1988).
I.~Antoniadis,
Phys.\ Lett.\ B {\bf 246}, 377 (1990);

\bibitem{defect}
K.~Akama,
Lect.\ Notes Phys.\  {\bf 176}, 267 (1982) [arXiv:hep-th/0001113];
V.~A.~Rubakov and M.~E.~Shaposhnikov,
Phys.\ Lett.\ B {\bf 125}, 139 (1983);
Phys.\ Lett.\ B {\bf 125}, 136 (1983); M.~Visser,
Phys.\ Lett.\ B {\bf 159}, 22 (1985) [arXiv:hep-th/9910093];
G.~W.~Gibbons and D.~L.~Wiltshire,
  Nucl.\ Phys.\  B {\bf 287} (1987) 717
  [arXiv:hep-th/0109093].

 \bibitem{RS} L.~Randall and R.~Sundrum,
Phys.\ Rev.\ Lett.\ {\bf 83}, 3370 (1999) [arXiv:hep-ph/9905221];
Phys.\ Rev.\ Lett.\ {\bf 83}, 4690 (1999) [arXiv:hep-th/9906064].

\bibitem{UED}
T.~Appelquist, H.~C.~Cheng and B.~A.~Dobrescu,
Phys.\ Rev.\ D {\bf 64}, 035002 (2001) [arXiv:hep-ph/0012100];
J.~L.~Feng, A.~Rajaraman and F.~Takayama,
Phys.\ Rev.\ D {\bf 68}, 085018 (2003) [arXiv:hep-ph/0307375].

\bibitem{reviews}
V.~A.~Rubakov,
Phys.\ Usp.\  {\bf 44}, 871 (2001) [Usp.\ Fiz.\ Nauk {\bf 171},
913 (2001)] [arXiv:hep-ph/0104152]; G.~Gabadadze,
[arXiv:hep-ph/0308112];
 C.~Csaki,
  [arXiv:hep-ph/0404096].

\bibitem{GRW} G.~F.~Giudice, R.~Rattazzi and J.~D.~Wells,
Nucl.\ Phys.\ B {\bf 544}, 3 (1999) [arXiv:hep-ph/9811291];
T.~Han, J.~D.~Lykken and R.~J.~Zhang,
Phys.\ Rev.\ D {\bf 59}, 105006 (1999) [arXiv:hep-ph/9811350];
R. Emparan, M. Masip and R. Rattazzi, Phys.\ Rev.\ D {\bf 65}, 064023 (2002)
[arXiv:hep-ph/0109287].

\bibitem{LandauII} See for instance L. Landau and E. Lifshitz Volume II, Section 73.

\bibitem{bremnr}
S.~Cullen and M.~Perelstein,
Phys.\ Rev.\ Lett.\  {\bf 83}, 268 (1999) [arXiv:hep-ph/9903422];
L.~J.~Hall and D.~R.~Smith,
Phys.\ Rev.\ D {\bf 60}, 085008 (1999) [arXiv:hep-ph/9904267];
V.~D.~Barger, T.~Han, C.~Kao and R.~J.~Zhang,
Phys.\ Lett.\ B {\bf 461}, 34 (1999) [arXiv:hep-ph/9905474];
C.~Hanhart, D.~R.~Phillips, S.~Reddy and M.~J.~Savage,
  Nucl.\ Phys.\  B {\bf 595}, 335 (2001)
  [arXiv:nucl-th/0007016];
  S.~Hannestad and G.~G.~Raffelt,
  Phys.\ Rev.\  D {\bf 67}, 125008 (2003)
  [Erratum-ibid.\  D {\bf 69}, 029901 (2004)]
  [arXiv:hep-ph/0304029];
V.~H.~Satheeshkumar and P.~K.~Suresh,
  JCAP {\bf 0806}, 011 (2008)
  [arXiv:0805.3429 [astro-ph]].

\bibitem{MPP}E.~A.~Mirabelli, M.~Perelstein and M.~E.~Peskin,
Phys.\ Rev.\ Lett.\  {\bf 82}, 2236 (1999) [arXiv:hep-ph/9811337];
J.~L.~Hewett,
Phys.\ Rev.\ Lett.\  {\bf 82}, 4765 (1999) [arXiv:hep-ph/9811356].

\bibitem{Dvergsnes:2002nc}
  E.~Dvergsnes, P.~Osland and N.~Ozturk,
  Phys.\ Rev.\  D {\bf 67}, 074003 (2003)
  [arXiv:hep-ph/0207221];
  T.~Buanes, E.~W.~Dvergsnes and P.~Osland,
  arXiv:hep-ph/0408063;
  Eur.\ Phys.\ J.\  C {\bf 35}, 555 (2004)
  [arXiv:hep-ph/0403267].
  E.~Dvergsnes, P.~Osland and N.~Ozturk,
  arXiv:hep-ph/0108029.
  X.~G.~Wu and Z.~Y.~Fang,
  Phys.\ Rev.\  D {\bf 78}, 094002 (2008)
  [arXiv:0810.3314 [hep-ph]].

\bibitem{gkst}
  D.~V.~Gal'tsov, G.~Kofinas, P.~Spirin and T.~N.~Tomaras, {\bf JHEP} 0905:074, 2009;
 arXiv:0903.3019 [hep-ph].


\bibitem{thooft} G. 't Hooft, Phys. Lett. B 198 (1987) 61;
I.J. Muzinich and M. Soldate, Phys. Rev. D37 (1988) 359;
D. Amati, M. Ciafaloni and G. Veneziano, Nucl. Phys. B 403 (1993) 707.

\bibitem{bere}
V. B. Berestetskii, E. M. Lifshitz and L. P. Pitaevskii,
Relativistic Quantum Theory , Part 1, Pergamon Press (1971)
\bibitem{GaGra}
  D.~V.~Galtsov and Yu.~V.~Grats,
  Teor.\ Mat.\ Fiz.\  {\bf 28}, 201 (1976).

\bibitem{minkd}
  B.~P.~Kosyakov,
  Theor.\ Math.\ Phys.\  {\bf 119}, 493 (1999)
  [Teor.\ Mat.\ Fiz.\  {\bf 119}, 119 (1999)]
  [arXiv:hep-th/0207217];
  D.~V.~Galtsov,
  Phys.\ Rev.\  D {\bf 66}, 025016 (2002)
  [arXiv:hep-th/0112110];
  P.~O.~Kazinski, S.~L.~Lyakhovich and A.~A.~Sharapov,
  Phys.\ Rev.\  D {\bf 66}, 025017 (2002)
  [arXiv:hep-th/0201046];
  D.~V.~Gal'tsov  and P.~A.~Spirin,
  Grav.\ Cosmol.\  {\bf 13} (2007) 241.

\bibitem{minkd1}
  V.~Cardoso, O.~J.~C.~Dias and J.~P.~S.~Lemos,
  Phys.\ Rev.\  D {\bf 67}, 064026 (2003)
  [arXiv:hep-th/0212168];
  M.~Gurses and O.~Sarioglu,
  Class.\ Quant.\ Grav.\  {\bf 19}, 4249 (2002)
  [Erratum-ibid.\  {\bf 20}, 1413 (2003)]
  [arXiv:gr-qc/0203097];
  B.~Koch and M.~Bleicher,
  JETP Lett.\  {\bf 87}, 75 (2008)
  [arXiv:hep-th/0512353];
  P.~Krotous and J.~Podolsky,
  Class.\ Quant.\ Grav.\  {\bf 23}, 1603 (2006)
  [arXiv:gr-qc/0602007];
  V.~Cardoso, M.~Cavaglia and J.~Q.~Guo,
  Phys.\ Rev.\  D {\bf 75}, 084020 (2007)
  [arXiv:hep-th/0702138];
  V.~Cardoso, O.~J.~C.~Dias and J.~P.~S.~Lemos,
  Phys.\ Rev.\  D {\bf 67}, 064026 (2003)
  [arXiv:hep-th/0212168];
B.P.~Kosyakov,  ``Introduction to the classical theory of
particles and fields'', Springer, 2007;
  A.~Mironov and A.~Morozov,
  Pisma Zh.\ Eksp.\ Teor.\ Fiz.\  {\bf 85}, 9 (2007)
  [JETP Lett.\  {\bf 85}, 6 (2007)]
  [arXiv:hep-ph/0612074];
  A.~Mironov and A.~Morozov,
  arXiv:0710.5676 [hep-th];
  A.~Mironov and A.~Morozov,
  arXiv:hep-th/0703097


\bibitem{gkst3} D. Galtsov, G. Kofinas, P. Spirin and T.N. Tomaras, Phys. Lett. {\bf B683} (2010) 183;
arXiv:0908.0675.

\bibitem{gkst4} D. Galtsov, G. Kofinas, P. Spirin and T.N. Tomaras, in preparation.

\bibitem{landauIII} L. Landau and E. Lifshitz, {Quantum Mechanics}

\bibitem{giudice} G. Giudice, R. Rattazzi and J. Wells, Nucl. Phys. {\bf B} 630 (2002) 293.

\bibitem{maxi}
H. Bethe, L. Maximon, Phys. Rev. {\bf 93}, 768 (1954).

\bibitem{GR}  I.S. Gradshteyn and I.M. Ryzhik , "Table of Integrals, Series and Products",
Academic Press 1965.

  \bibitem{Proudn}
Proudnikov A.P. "Integrals and series", vol.1,2 [in russian],
 Nauka, Moscow, 1981

\end{thebibliography}
\end{document}